\documentclass[a4paper,superscriptaddress,twocolumn,nofootinbib,longbibliography]{aastex631}

\usepackage{aas_macros}
\usepackage{acro}
\acsetup{patch/longtable=false}
\usepackage{graphicx}
\usepackage{dcolumn}
\usepackage{bm}
\usepackage{comment}
\usepackage{amsmath}
\usepackage{float}
\usepackage{lipsum}
\usepackage{multirow}
\usepackage{makecell}
\usepackage{pifont}
\usepackage{color}
\usepackage{gensymb}
\usepackage{subfigure}
\usepackage{times}
\usepackage{textcomp}
\usepackage{tabularx}
\usepackage{latexsym}
\usepackage{mathtools}
\usepackage{url}
\usepackage[utf8]{inputenc}
\usepackage[normalem]{ulem}
\usepackage{physics}
\usepackage{subfigure}
\usepackage{soul}
\usepackage{hyperref}
\hypersetup{colorlinks,linkcolor={blue},citecolor={blue},urlcolor={red}}  
\setcitestyle{authoryear,open={},close={}}
\setcitestyle{open={(},close={)}}
\usepackage{color}
\definecolor{mygreen}{rgb}{0.1,0.8,0.1}
\usepackage{algorithm,algpseudocode}
\newcommand{\cmark}{\ding{51}}
\newcommand{\xmark}{\ding{55}}
\algnewcommand{\Inputs}[1]{%
  \State \textbf{Inputs:}
  \Statex \hspace*{\algorithmicindent}\parbox[t]{.8\linewidth}{\raggedright #1}
}
\algnewcommand{\Initialize}[1]{%
  \State \textbf{Initialize:}
  \Statex \hspace*{\algorithmicindent}\parbox[t]{.8\linewidth}{\raggedright #1}
}

\newcommand{\jj}[1]{\textcolor{blue}{[JJ: #1]}}

\begin{document}
\title{The impact of strong lensing on Hubble constant measurements with gravitational-wave dark sirens}

\author{Eungwang~\surname{Seo}}
\thanks{e.seo.1@research.gla.ac.uk}
\affiliation{SUPA, School of Physics and Astronomy, University of Glasgow, Glasgow G12 8QQ, United Kingdom}

\author{Kyungmin~\surname{Kim}}
\thanks{kkim@kasi.re.kr}
\affiliation{Korea Astronomy and Space Science Institute, 776 Daedeokdae-ro, Yuseong-gu, Daejeon 34055, Republic of Korea}

\author{Zhuotao~\surname{Li}}
\affiliation{SUPA, School of Physics and Astronomy, University of Glasgow, Glasgow G12 8QQ, United Kingdom}

\author{Justin Janquart}
\affiliation{Center for Cosmology, Particle Physics and Phenomenology - CP3, Universit\'e Catholique de Louvain, Louvain-La-Neuve, B-1348, Belgium}
\affiliation{Royal Observatory of Belgium, Avenue Circulaire, 3, 1180 Uccle, Belgium}

\author{Rachel~\surname{Gray}}
\affiliation{SUPA, School of Physics and Astronomy, University of Glasgow, Glasgow G12 8QQ, United Kingdom}

\author{Martin~\surname{Hendry}}
\affiliation{SUPA, School of Physics and Astronomy, University of Glasgow, Glasgow G12 8QQ, United Kingdom}

\begin{abstract}
The disagreement between early and late Universe electromagnetic measurements of the Hubble constant, $H_0$, known as the Hubble tension, highlights the need for independent and complementary probes.
Gravitational-wave events have recently emerged as such a probe for constraining cosmological parameters. 
$H_{0}$ inference using these events relies on sky localization and luminosity distance estimates, both of which can be significantly improved for strongly lensed events with appropriate lens modeling. 
In this context, we propose utilizing strong lensing of dark sirens, gravitational-wave events without identified electromagnetic counterparts, in combination with strong lensing of galaxies as a novel method for measuring $H_0$.
The constant is inferred from the luminosity distances of these lensed dark sirens and the redshifts of their host galaxies, combining information from individual events to obtain statistically stronger constraints when multiple events are available.
We adopt a simulated galaxy catalog, \texttt{MICECATv2}, as the basis for simulating strong lensing of galaxies and to provide the redshift information of host galaxy candidates required to infer $H_0$. 
We also examine the impact of galaxy catalog incompleteness on the resulting $H_0$ inference.
Our results demonstrate that using only 8 strongly lensed dark sirens, analyzed with a dedicated galaxy-galaxy lensing catalog, can improve the precision of $H_{0}$ by roughly 50\% compared to 250 unlensed events. 
\end{abstract}

\section{Introduction}

In modern cosmology, the Hubble constant, $H_0$, plays a central role in understanding the evolution of the Universe. 
Despite its significance, the precise value of $H_{0}$ remains under debate due to the persistent discrepancy, the so-called Hubble tension~\citep{di2021realm}, between early-Universe measurements, such as those inferred from the cosmic microwave background~\citep{Planck:2018vyg} reporting $H_0=67.4 \pm 0.5~\mathrm{km}~\mathrm{s}^{-1}~\mathrm{Mpc}^{-1}$, and late-Universe measurements based on the distance ladder using Cepheid variable stars and Type Ia supernovae~\citep{2022ApJ...934L...7R} reporting $H_0=72.3 \pm 1.4~\mathrm{km}~\mathrm{s}^{-1}~\mathrm{Mpc}^{-1}$. This tension has sparked considerable interest in independent and novel methods for determining $H_{0}$.

Over the past decade, the era of gravitational-wave (GW) astronomy has flourished with the detection of more than 200 GW signals~\citep{LIGOScientific:2025slb} by the ground-based LIGO–Virgo–KAGRA (LVK) detector network~\citep{LIGOScientific:2014pky,VIRGO:2014yos,KAGRA:2018plz}.
To date, the sources of all confirmed GW events have been consistent with the coalescence of two relativistic compact objects, such as black holes (BHs) or neutron stars (NSs)~\citep{LIGOScientific:2025slb}.
The inferences drawn from the GW signals have impacted multiple areas of astrophysics: results to date have clarified the statistical distributions of compact-object masses and spins~\citep{LIGOScientific:2025pvj}, provide strong evidence for the existence of intermediate-mass BHs~\citep{LIGOScientific:2025rsn}, place constraints on the equation of state for dense matter in NS interiors~\citep{LIGOScientific:2018cki}, and enable precision tests of general relativity in the highly dynamical, strong-field regime~\citep{abbott2025tests}.

With its current sensitivity, the LVK network is capable of detecting GWs from sources at luminosity distances of a few gigaparsecs~\citep{LIGOScientific:2025slb}, thereby extending GW observations to cosmological scales. 
Such reach enables GW observations to probe cosmological models and constrain fundamental parameters~\citep{Zheng:2022gfi, Jin:2025dvf}. 
Since the luminosity distance can be directly inferred from the waveform amplitude of compact binary coalescences (CBCs), without reference to local calibrators, these sources are referred to as standard sirens~\citep{schutz1986determining,holz2005using}.
If the redshifts of the sources can be determined separately, GW events provide a measurement of $H_0$ in a completely new way.
In particular, events with identified electromagnetic (EM) counterparts, so-called bright sirens, allow for a direct determination of the source redshift~\citep{LIGOScientific:2017adf,2025RSPTA.38340134S}.
In contrast, the redshift must be inferred statistically for dark sirens, for which EM counterparts are absent~\citep{DES:2019ccw,mukherjee2021accurate,LIGOScientific:2019zcs,gray2023joint,2023RNAAS...7..250B,2024MNRAS.535..961B}). Several statistical methodologies have been developed.

One method exploits characteristic features in the mass distributions of neutron stars and black holes to partially break the degeneracy between the intrinsic source mass and cosmological redshift, which we refer to as the spectral siren method~\citep{ezquiaga2022spectral}.
Cross-correlation techniques instead constrain $H_0$ by comparing the large-scale angular distribution of galaxies in a catalog with the luminosity distances of GW events~\citep{mukherjee2024cross,afroz2024prospect,de2025first}.
In the galaxy-catalog-based dark siren approach~\citep{PhysRevD.101.122001,gray2023joint}, which forms the basis of our work, each GW event is statistically associated with multiple candidate host galaxies within its sky localization region.
By weighting these galaxies according to their spatial probability and luminosity, one constructs a redshift probability distribution for the GW source, enabling hierarchical inference of $H_{0}$.

For the galaxy-catalog-based method, the key to improving constraints on $H_{0}$ lies in two aspects: accurately localizing the source to limit the number of potential host galaxies, and obtaining precise luminosity distance measurements.
In this context, strongly lensed GWs (SLGWs) provide more informative constraints than unlensed events, improving sky localization~\citep{Hannuksela:2020xor, uronen2025finding} and luminosity distance estimates, both essential for host-galaxy identification in the dark-siren method, while also providing lensing information that can enhance cosmological constraints on $H_0$.
In particular, we have recently demonstrated that the luminosity distance inferred from simulated signals of SLGW can have a smaller uncertainty than that inferred from unlensed sources, due to the magnification and multiple images providing redundancy in measurement that can result in more effectively constraining the source parameters~\citep{kim2024gravitational}.

Although no unambiguous detection of lensed GW signals has yet been identified in events reported in the Gravitational-Wave Transient Catalogs to date~\citep{hannuksela2019search,abbott2021search,LIGOScientific:2023bwz,janquart2023follow,ligo2025gwtc}, the LVK detector network is expected to detect SLGWs at a rate of $\mathcal{O}(1)~\mathrm{Gpc}^{-3} \mathrm{yr}^{-1}$ at their design sensitivities~\citep{2018MNRAS.476.2220L,2018PhRvD..97b3012N,2018MNRAS.480.3842O,2021MNRAS.506.3751M,2021ApJ...921..154W,xu2022please,LIGOScientific:2023bwz}, offering advantages for $H_0$ inference similar to those obtained from time-delay measurements of strongly lensed quasars~\citep{2020MNRAS.498.1420W, Birrer:2022chj, 2024A&A...689A.168W}.

As more SLGW will be detected not only in the upcoming observing runs of the LVK network~\citep{LIGOScientific:2023bwz,2025RSPTA.38340134S} but also with next-generation detectors~\citep{Zheng:2022gfi,Abac:2025saz}, such as Einstein Telescope~\citep{2010CQGra..27s4002P}, it will be crucial to perform a joint analysis incorporating population models of both the sources and the lenses. Furthermore, unlensed dark sirens should not be excluded from the analysis; combining both lensed and unlensed GW events in a hierarchical framework can significantly enhance the precision and accuracy of the inferred $H_0$. In this regard, tools such as \texttt{gwcosmo}~\citep{PhysRevD.101.122001,gray2022pixelated,gray2023joint} can provide a flexible framework for incorporating population information and performing Bayesian inference on cosmological parameters from both lensed and unlensed GW sources.

In this work, we investigate the impact of SLGWs on $H_{0}$ inference, highlighting that lensing provides complementary information that strengthens cosmological constraints relative to unlensed dark sirens.
Two observational scenarios are considered. 
In the first, a single set of lensed dark sirens is detected. 
Once the host galaxy is uniquely identified through lensing observables and cross-correlation with galaxy catalogs, the Hubble-Lemaître law can be directly applied to infer $H_{0}$.
In the second scenario, multiple sets of lensed dark sirens are detected, where a hierarchical framework is required to combine the dark siren events consistently.
The framework leverages comparisons with galaxy–galaxy lensing catalogs to construct more informative redshift maps and thereby improve the precision of the inferred $H_{0}$.
Furthermore, we quantify the systematic biases in $H_{0}$ inference that can arise if lensing effects are neglected or misinterpreted.

The main results show that a single quadruply lensed dark siren drawn from a realistic lensing population model can constrain $H_{0}$ to a precision of up to $\sim 20\%$.
Utilizing multiple sets of SLGWs alone improves the precision of the inferred $H_{0}$ by approximately 50\% relative to a scenario with a few hundred unlensed dark-siren events.
Regarding potential biases due to lensing misinterpretation, these effects are highly asymmetric.
If a small number of unlensed events are incorrectly analyzed as lensed, the recovered $H_{0}$ can exhibit a substantial systematic bias.
In contrast, when a few lensed events are treated as unlensed, the inferred $H_{0}$ shifts only by $\Delta H_{0} \sim \mathcal{O}(1)~\mathrm{km~s^{-1}~Mpc^{-1}}$.
This behavior highlights that careful lensing identification is crucial to avoid large biases, while failing to discern lensed events mainly affects statistical precision rather than causing significant systematic errors.

This paper is organized as follows: In Section~\ref{sec:SL}, we provide a brief overview of strong gravitational lensing.
In Section~\ref{sec:mockdata}, we describe the method adopted in this work to prepare the mock data of galaxies and dark sirens.
In Section~\ref{sec:H0inference}, we elaborate on the results of inferring $H_0$ obtainable in two ways, using single dark sirens (Section~\ref{subsec:HLlaw}) and multiple dark sirens (Section~\ref{subsec:gwcosmo}).
We then outline challenges in the SLGW-based measurement of $H_0$ due to considerable lensing biases in Section~\ref{sec:challenges} and discuss several key points that will be critical and required in considering a realistic observational scenario in Section~\ref{sec:discussion}.

Throughout this paper, the fiducial cosmology is assumed to be a flat $\Lambda$CDM model with $\Omega_{m}=0.25$, $\Omega_{b}=0.044$, $\Omega_{\Lambda}=0.7$, and $H_{0}=70~\rm{km,s^{-1},Mpc^{-1}}$.

\section{Strong gravitational lensing}\label{sec:SL}
When GWs or EM waves propagate near a massive object, they are deflected by the gravitational potential of the object~\citep{bartelmann2010gravitational}.
For most EM radiation, except at extremely long wavelengths, the wavelength is much shorter than the characteristic scale of lensing objects (e.g., star clusters, galaxies, and galaxy clusters).
In this regime, the geometric optics approximation is valid, and the original wave is split into multiple distinct images that are (de-)magnified and delayed in arrival time.
Each image is magnified by a factor $\mu$ and experiences a time delay arising from both the geometric path difference and the gravitational potential well (the so-called Shapiro time delay)~\citep{shapiro1964fourth,schneider2006gravitational}.

Although GWs are characterized by much longer wavelengths than those of EM waves, the same geometrical optics approximation remains valid for CBCs detectable by the LVK network when lensed by galaxies or galaxy clusters. 
In this case, the GW wavelength is smaller than, or at most comparable to, the Schwarzschild radius of the lens.
Analogous to the EM case, a GW signal ($h$) passing by near a galactic- or cluster-scale lens is split into multiple distinct lensed signals ($h_{l}$), each characterized by different amplitudes, arrival times, and phases~\citep{wang1996gravitational,takahashi2003wave,dai2017waveforms}.
In the frequency domain, the relationship between the unlensed waveform, $h(f)$, and the lensed waveform, $h_{l}(f)$, is described by the amplification factor, $F(f)$, which is defined as
\begin{equation}
\label{eq:lensedwaveform}
F(f,\boldsymbol{\theta_{l}}) \equiv \frac{h_{l}(f,\boldsymbol{\theta_{l}})}{h(f)} =\sum_{i}|\mu_{i}|^{0.5}  \exp{i\pi\left(2ft_{i} - n_{i}\right)} ,
\end{equation}
where $\boldsymbol{\theta_{l}}$ denotes lens parameters, and $\mu_{i}$, $t_{i}$, and $n_{i}$ are the magnification factor, time delay, and Morse number of the $i^{\rm th}$ lensed signal, respectively~\citep{takahashi2003wave}.

Assuming that a galaxy hosting a GW source is lensed by a foreground galaxy, the GW signal is likewise lensed by the same galaxy, producing multiple lensed signals in correspondence with the lensed images of the host galaxy.
The image positions of the lensed GW signals on the lens plane are expected to lie within the extent of the corresponding lensed host-galaxy images.
Hence, the magnification factors evaluated at the GW image positions are identical to those of galaxy light at the same locations.
Owing to the extended nature of the galaxy images, the magnification can vary across the image plane. Specifically, for sources in the nearby Universe, the finite angular size of the lensed host-galaxy images may become non-negligible, leading to differences between the magnification factors and time delays of the lensed GW signals and those of the extended source-galaxy images.
However, such variations are typically negligible for distant extragalactic sources.

The $i^{\rm th}$ lensed GW signal and its corresponding lensed host-galaxy image therefore share the same magnification factor, $\mu_{i}$.
In galaxy surveys, the magnification factor modifies the observed flux of the host galaxy and is reported in terms of an apparent magnitude.
In contrast, for GW observations, the magnification factor affects the inferred luminosity distance through the amplitude of the measured waveform.
Consequently, the apparent magnitude of the $i^{\rm th}$ lensed source-galaxy image, $m'_{i}$ , and the effective luminosity distance inferred for the $i^{\rm th}$ lensed GW signal, $d^{\rm eff}_{\mathrm{L},i}$, can be written as
\begin{equation}
\label{eq:lensed_observables}
\begin{aligned}
m'_i &= m - 2.5 \log_{10}(\mu_i), \\
d^{\rm eff}_{\mathrm{L},i} &= \frac{d_{\rm L}}{\sqrt{\mu_i}},
\end{aligned}
\end{equation}
where $m$ and $d_{\rm L}$ denote the unlensed apparent magnitude of the host galaxy and the true luminosity distance to the GW source, respectively.
If $\mu > 1$, lensing increases the observed flux of the source, rendering the lensed galaxy image brighter and hence lowering its apparent magnitude.
For GWs, the observed strain amplitude scales inversely with the luminosity distance, such that lensing leads to a smaller inferred luminosity distance.

Additionally, the relative magnification factor ($\mu_{\rm rel, EM}$) between the $i^{\mathrm{th}}$ and $j^{\mathrm{th}}$ lensed galaxy images is given by
\begin{equation}
\label{eq:em_mu_rel}
    \mu_{\rm rel, EM} \equiv \frac{\mu_{j}}{\mu_{i}} = 10^{-0.4 (m'_{j}-m'_{i})}.
\end{equation}
Similar to the EM case, the relative magnification factor ($\mu_{\rm rel, GW}$) between $i^{\mathrm{th}}$ and $j^{\mathrm{th}}$ lensed GW signals is defined as
\begin{equation}
\label{eq:gw_mu_rel}
    \mu_{\rm rel, GW} \equiv \frac{\mu_{j}}{\mu_{i}} = \left(\frac{d^{\rm eff}_{\rm{L},\it i}}{d^{\rm eff}_{\rm{L},\it j}}\right)^{2}.
\end{equation}
If the magnification is assumed to be approximately constant across the lensed galaxy image, $\mu_{\rm rel, EM}$ and $\mu_{\rm rel, GW}$ in Eqs.~\eqref{eq:em_mu_rel} and~\eqref{eq:gw_mu_rel}, respectively, are identical.

In addition to magnification, lensing introduces relative time delays between multiple images.
For transient signals, including GW signals from CBCs and EM transients such as supernovae or gamma-ray bursts, these time delays can be measured directly.
For instance, in the bright-siren case, the time delay between the arrival times of the $i^{\rm th}$ and $j^{\rm th}$ lensed EM or GW signals, denoted $t_i$ and $t_j$, are given by
\begin{equation}
\label{eq:transient_delta_t}
    \Delta t_{\rm EM} = \Delta t_{\rm GW} \equiv |t_{i} - t_{j}|.
\end{equation}
For galaxy–galaxy strong lensing in EM observations, however, the time delays between multiple images are generally not measurable.
Instead, they are obtained using the Fermat potentials, $\tau_i$ and $\tau_j$, evaluated at the corresponding image positions using an assumed lens model. 
The relative time delay between the $i^{\rm th}$ and $j^{\rm th}$ images is then given by
\begin{equation}
\label{eq:em_delta_t}
    \Delta t_{\rm EM} \equiv |t_{i} - t_{j}|=\frac{1+z_{L}}{c} \frac{D_{L}D_{S}}{D_{LS}} (|\tau_{i} -\tau_{j}|),
\end{equation}
where $z_{L}$ is the lens redshift, $D_{L}$,$D_{S}$, and $D_{LS}$ are the angular diameter distances from the observer to the lens, from the lens to the source, and from the observer to the source, respectively.
When multiple images are identified in the galaxy catalog, $\mu_{\rm rel, EM}$ and $\Delta t_{\rm EM}$ can be computed for each image pair in the lensing system.
The quantities, $\mu_{\rm rel}$ and $\Delta t$, are referred to as lensing observables, since the individual magnification factors and the absolute delays relative to the unlensed source cannot be directly measured.
\section{Mock data}\label{sec:mockdata}
\begin{figure*}[ht]
    \centering
\includegraphics[width=0.8\linewidth]{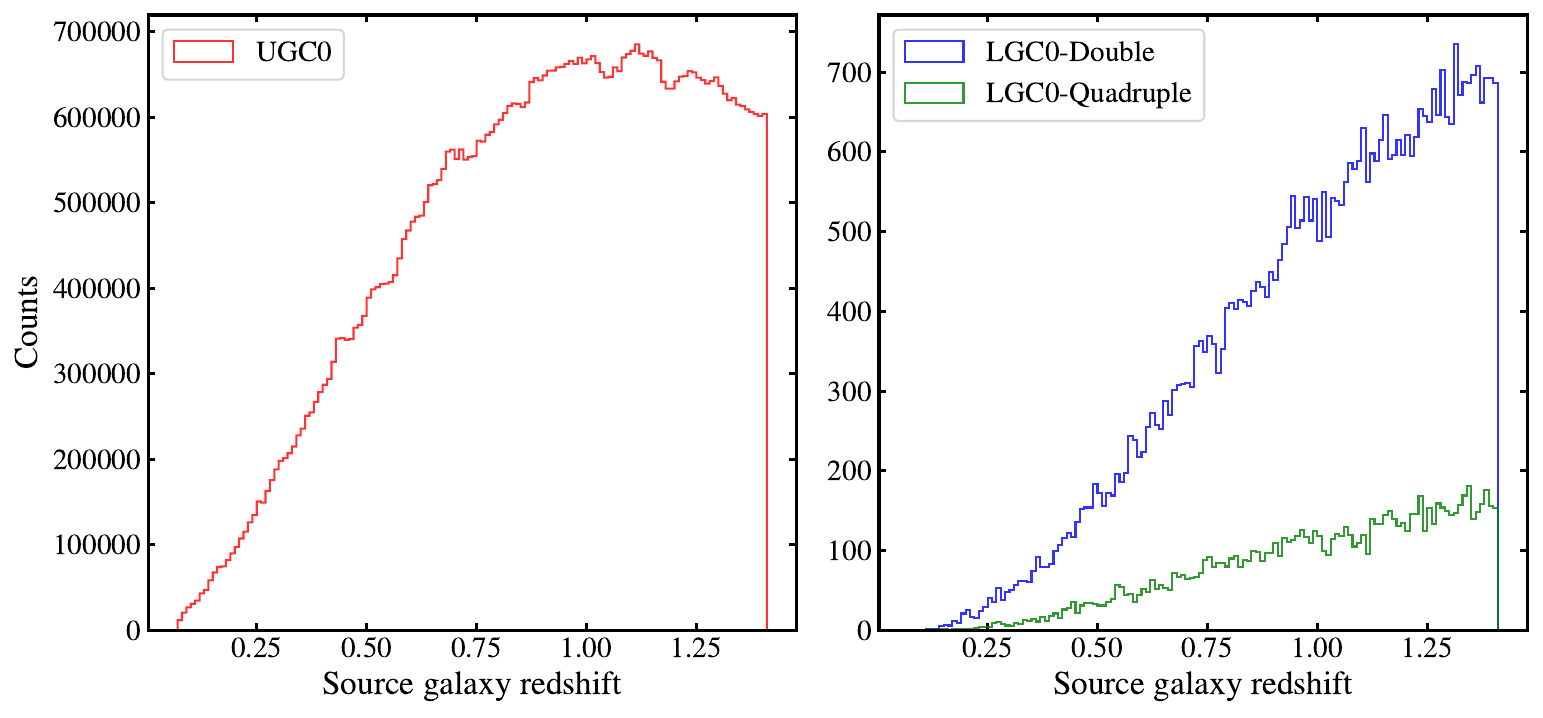}
    \caption{
    \textit{Left panel}: The redshift distribution of galaxies in the unlensed catalog, UGC0 (red), which contains approximately 64 million galaxies.
    \textit{Right panel}: The redshift distributions of galaxies in the lensed catalog, LGC0. The number of quadruple-image systems (green) is smaller than that of double-image systems (blue) because the lensing cross-section for quadruple-image formation is much smaller than that for double-image formation.}
    \label{fig:gal_dist}
\end{figure*}
\subsection{Galaxy catalog}~\label{subsec:gc}
We adopt the MICE-Grand Challenge light-cone halo and galaxy catalog (\texttt{MICECATv2})~\citep{Fosalba2014mice3,Carretero2014mice4,Hoffmann2014mice5,fosalba2015mice,crocce2015mice} for our galaxy distribution, while separately investigating how catalog incompleteness affects the inference of $H_{0}$.
Galaxies are populated in the simulation volume, which covers approximately 5000~deg$^2$ of the sky, over the redshift range $0.07 < z < 1.42$.
The catalog provides both luminosities and apparent magnitudes in multiple photometric bands, as well as true and observer-frame redshifts.
We select the $r$-band for our analysis since \texttt{MICECATv2} is calibrated using the $r$-band Schechter function from \cite{blanton2003galaxy}, with galaxy evolution applied at higher redshift~\citep{carretero2015algorithm}.
Although the completeness estimates were originally derived using $i$-band data, we adopt the $r$-band to ensure a direct and consistent application of the Schechter parameters for incompleteness correction.
Here, completeness is defined as the fraction of galaxies that are observable and included in the catalog.

For computational efficiency in \texttt{gwcosmo}, we use a randomly selected one-eighth subset of the full \texttt{MICECATv2} catalog, which contains \(\sim 6.2\times10^{7}\)  galaxies\footnote{The number density of galaxies is 0.00146 $\rm{Mpc}^{-3}$ (the original one was 0.0117 $\rm{Mpc}^{-3}$), which is more similar to the galaxy number density of real galaxy catalogs, such as GLADE+~\citep{dalya2022glade+}, used in GW cosmology analyses with dark sirens.}, all with absolute magnitudes $M_{r} < -20$, and assume this subset to be 100\% complete down to an apparent magnitude $m_{r} < 24$ at $z=1.42$~\citep{carretero2015algorithm}.
We impose the absolute-magnitude cut to the catalog to obtain an approximately uniform number density of galaxies in comoving volume, ensuring that the catalog is effectively volume-limited, which prevents an overabundance of faint, low-redshift galaxies while avoiding significant bias toward bright galaxies.
Note that we exclude galaxy clusters to account for only galaxy-galaxy strong lensing.
Hereafter, this complete catalog is referred to as the Unlensed Galaxy Catalog (UGC0).

Within the landscape of galaxy surveys, certain catalogs are restricted to lensed galaxies, such as the SLACS survey~\citep{bolton2006sloan,bolton2008sloan}, which targets systems where strong lensing has been confirmed.
Complementary efforts have assembled dedicated catalogs of strong gravitational lenses that facilitate the identification of host galaxies associated with lensed transient events, such as \texttt{lenscat}~\citep{vujeva2025lenscat}.
In this context, we also simulate a catalog of lensed galaxies derived from UGC0.
Relative to the unlensed case, the lensed galaxy catalog contains far fewer potential host-galaxy candidates per unit sky area within the GW localization region, thereby mitigating the source-identification uncertainty in the statistical $H_{0}$ inference with \texttt{gwcosmo}.
To simulate gravitational lensing systems, we assume that all galaxies in our mock catalog follow singular isothermal ellipsoid (SIE) mass profiles, characterized by their axis ratio ($q$), orientation angle ($\phi_{e}$), and velocity dispersion ($\sigma_{v}$)~\citep{kormann1994isothermal}, which are called lens parameters.

Since \texttt{MICECATv2} does not directly provide these lens parameters, we model their underlying distributions.
We adopt the same functional forms as those used in \citet{xu2022please}. However, rather than fixing the galaxy number density at a specific redshift in the Schechter function, we compute it independently at each redshift using the \texttt{MICECATv2} data and sample $\sigma_{v}$ accordingly.
For the shape parameters, we assume a uniform distribution for $\phi_{e}$ and a Rayleigh distribution for $q$, extending the lower limit to $q_{\rm min}=0.1$ in order to include highly elliptical galaxies.

The number of strongly lensed galaxies, $N_{\rm Lgal}$, in a given galaxy catalog is determined by the optical depth, $\tau_{\rm gal}$, which is given by
\begin{equation}
\begin{aligned}
    N_{\rm Lgal} &= N_{\rm gal} \times \tau_{\rm gal}\\
    &= N_{\rm gal} \times \sum^{N_{\rm gal}}_{i} \int \tau(z_{i})P(z_{i}) dz_{i},
\end{aligned}
\end{equation}
where $\tau(z_i)$ is the optical depth for the $i^{\rm th}$ galaxy at redshift $z_i$, and $P(z_i)$ represents the probability that a galaxy from the catalog resides at redshift $z_i$.
Hence, $\tau_{\rm gal}$ corresponds to the combined optical depth for all galaxies in the given universe (see Appendix~\ref{app:optical_depth} for further details).

Since we adopt an SIE lens model, only two or four images of a source galaxy can be produced.
The lensed galaxy catalog contains \(\sim 4.6\times10^{4}\)  double-image systems and \(\sim 1.0\times10^{4}\)  quadruple-image systems.
Hereafter, the complete lensed galaxy catalog, containing both double- and quadruple-image systems generated from the UGC0, is referred to as the Lensed Galaxy Catalog (LGC0).
Figure~\ref{fig:gal_dist} shows the redshift distribution of galaxies in the UGC0, along with the distributions of double- and quadruple-image systems in the LGC0.
The distribution of lensed galaxies peaks at high redshift, which is expected, as galaxies at higher redshift have a higher probability of encountering a foreground lensing galaxy.

In practice, galaxy catalogs are incomplete and do not contain all potential host galaxies of GW sources.
This incompleteness must be taken into account when inferring the Hubble constant using dark sirens and galaxy catalogs.
To mimic observational limitations, we impose an apparent magnitude cut, $m_{\rm th}=21.87$, removing galaxies fainter than this threshold from both UGC0 and LGC0.
The resulting incomplete catalogs, hereafter referred to as UGC1 and LGC1, will be discussed in Sec.~\ref{subsec:gwcosmo}.
\subsection{Dark sirens}
\label{subsec:GWsignal}
For dark-siren applications, BBH populations are simulated using the prior ranges of source parameters and hyperparameters inferred from the LIGO-Virgo-KAGRA GWTC-3 population analysis \citep{abbott2023population}. 
Specifically, we adopt the \textsc{PowerLaw+Peak} model for the primary component mass ($m_{1}$) with a power-law slope $\alpha = 3.78$ over the mass range $[5,112]M_{\odot}$, together with a Gaussian component characterized by a mean $\mu_{g}=32.27M_{\odot}$ and a standard deviation $\sigma_{g}=3.88M_{\odot}$.
The relative contribution of the Gaussian component is set by a mixing fraction $\lambda_{g}=0.03$.
The secondary component mass ($m_{2}$) is modeled using a truncated power-law distribution with slope $\beta=0.81$ over the range $[5,m_{1}]M_{\odot}$.
For the redshift distribution and merger rate of BBHs, we adopt the star formation rate model of \citet{madau2014cosmic}, combined with the detector-frame merger-rate parameterization of \citet{callister2020shouts}.
We adopt a local merger rate of $R_{0}=20~\rm{Gpc^{-3}yrs^{-1}}$, redshift turning point of $z_{p}=2.47$, and and power-law indices $\gamma=4.59$ and $k=2.86$.
The observation time is fixed to $T_{\rm obs} = 2.26~\mathrm{yrs}$, consistent with the expected duration of the O4 observation run.
For the other BBH parameters, including the spin parameters, we assume uniform distributions.
Further details of the population models can be found in Appendix~\ref{app:populations}.

Each simulated BBH is assigned to a galaxy in the UGC0 catalog following a physically motivated weighting scheme, in which a galaxy’s likelihood of hosting a BBH depends on its star formation rate, stellar mass, metallicity, and the delay time between binary formation and merger~\citep{vijaykumar2024inferring,li2025using}.
A weighting scheme that explicitly incorporates all these factors can improve the convergence of the inferred $H_0$ by better matching the true BBH host distribution.
Conversely, an inaccurate or oversimplified prescription may introduce systematic biases in the $H_0$ posterior~\citep{hanselman2025gravitational,perna2025investigating}.
Rather than explicitly modeling the full set of host-galaxy properties, we adopt a simple luminosity-based weighting scheme as a practical approximation, assigning higher hosting probabilities to brighter galaxies.

The unlensed GW signals are generated from the BBHs using the \textsc{IMRPhenomXPHM-SpinTaylor} waveform model~\citep{pratten2021computationally,PhysRevD.111.104019} and injected into a detector network consisting of LIGO–Livingston, LIGO–Hanford, and Virgo at their expected O4 sensitivities~\citep{abbott2020prospects}.
We only consider signals detected in all three detectors with a network signal-to-noise ratio (S/N) above 8 ($\rho_{\rm net} > 8$).

In addition to the unlensed GW signals, SLGW signals are generated using the aforementioned unlensed signals and the galaxies in the LGC0 using Eq.~\eqref{eq:lensedwaveform}.
To calculate the number of lensed GW signals ($N_{\rm lgw}$), a procedure similar to that used for constructing the LGC0 is applied; the optical depth for BBHs ($\tau_{\rm BBH}$) is obtained considering the galaxies in the UGC0, giving
\begin{equation}
    N_{\rm lgw} = N_{\rm gw} \times \tau_{\rm BBH}.
\end{equation}
Note that $\tau_{\rm BBH}$ is the combined optical depth for all BBHs, which is computed using the full unlensed galaxy catalog, UGC0.

In practice, it is prohibitively difficult to compute the dimensionless impact parameters ($y$) for each galaxy–galaxy pair, taking into account the angular diameter distances between sources, lenses, and observers, as well as their sky positions, and then to construct each strong lensing system.
Therefore, we randomly select $N_{\rm lgw}$ source–lens galaxy pairs from LGC0, where each source galaxy is assumed to host a BBH and is lensed by its paired lens galaxy.
Consequently, arbitrary $y$ values that result in strong lensing pairs are assigned, drawn from a uniform distribution over the range $y = [0.01,1.2]$, where $1.2$ is the maximum value of $y$ for which strong lensing can occur given the adopted axis-ratio range of the SIE model.
Table~\ref{tab:params} summarizes the assumed distributions for representative source and lens parameters.

Among the simulated lensed GW signals, only those satisfying $\rho_{\rm net} > 8$ after processing with the standard GW search pipeline, using the same detector configuration adopted for the unlensed case, are regarded as detected\footnote{In a realistic detection, thresholds on the astrophysical probability ($p_{\rm astro}$) and false-alarm rate (FAR) are applied to ensure that detected signals are likely of astrophysical origin and not noise. 
In this work, we assume that all signals satisfy these criteria.}.
In a realistic observational scenario, one cannot know whether detected signals are lensed or not at the time of detection.
Candidate lensed pairs (or higher-order multiplets) must therefore first be identified by low-latency lensing search pipelines.
For example posterior-overlap tests select events with mutually consistent intrinsic parameters (e.g., component masses and spins) and sky localizations, while allowing for differences in arrival times and observed strain amplitudes, and assign ranking statistics that quantify the degree of consistency under the lensing hypothesis~\citep{haris2018identifying}.
Joint parameter estimation (JPE) is subsequently performed for high-ranking candidates.
In this step, the lensed evidence is computed by jointly analyzing the corresponding data segments under the lensing hypothesis, thereby assessing whether the candidate multiplet is coherently described as multiple lensed signals of the same GW source.

In this work, the true associations between lensed signals are known by construction, and we perform JPE directly on the correct lensed signal sets.
Only those systems for which the lensing hypothesis is statistically supported by the JPE analysis are retained for the subsequent cosmological inference.

\begin{table}[t]
    \caption{Assumed probability distributions for component masses ($m_1$, $m_2$) of the source of dark sirens and the main lens parameters, including the axis-ratio ($q$) and the velocity dispersion ($\sigma_{v}$), and dimensionless impact parameter ($y$), employed in simulating strongly lensed GW signals.
    The models for masses and lens parameters, except for $y$, follow on~\citet{abbott2023population} and~\citet{xu2022please}, respectively.
    For $y$, we impose an upper limit of 1.2, corresponding to the maximum value that can produce more than one lensed image. 
    Details on the sampling of each parameter are provided in Appendix~\ref{app:populations}.}
\begin{tabular*}{\linewidth}
{@{\extracolsep{\fill}} llll}
\toprule
\textbf{Parameter} &
\textbf{Distribution model} & \textbf{Range} \\
\hline
$m_{1}$ & \texttt{Power Law + Peak} & [5, 112]$M_{\odot}$  \\
$m_{2}$ & \texttt{Truncated Power Law} & [5, $m_{1}$]$M_{\odot}$\\
$q$ & \texttt{Rayleigh} & [0.1, 1.0]\\
$\sigma_{v}$ & \texttt{Schechter} & [50,500]$\rm km/s$\\
$y$ & \texttt{Uniform} & [0.01, 1.2]\\
        \hline
    \end{tabular*}
\label{tab:params}
\end{table}
\section{Estimating the Hubble constant}\label{sec:H0inference}
\subsection{Single lensed dark siren and the Hubble-Lema\^{i}tre law }\label{subsec:HLlaw}
Assuming the detection of a single lensed dark siren, we can obtain improved sky localization compared to the unlensed case~\citep{Hannuksela:2020xor,uronen2025finding} as well as more precise measurements of the luminosity distance ($d_{L}$)~\citep{kim2024gravitational}.
Within the more tightly constrained sky area, candidate lens galaxies may be identifiable, but whether the lensed images are detectable also depends on the lens mass, luminosity, and distance.

\begin{figure*}[t]
    \centering
\includegraphics[width=0.8\linewidth]{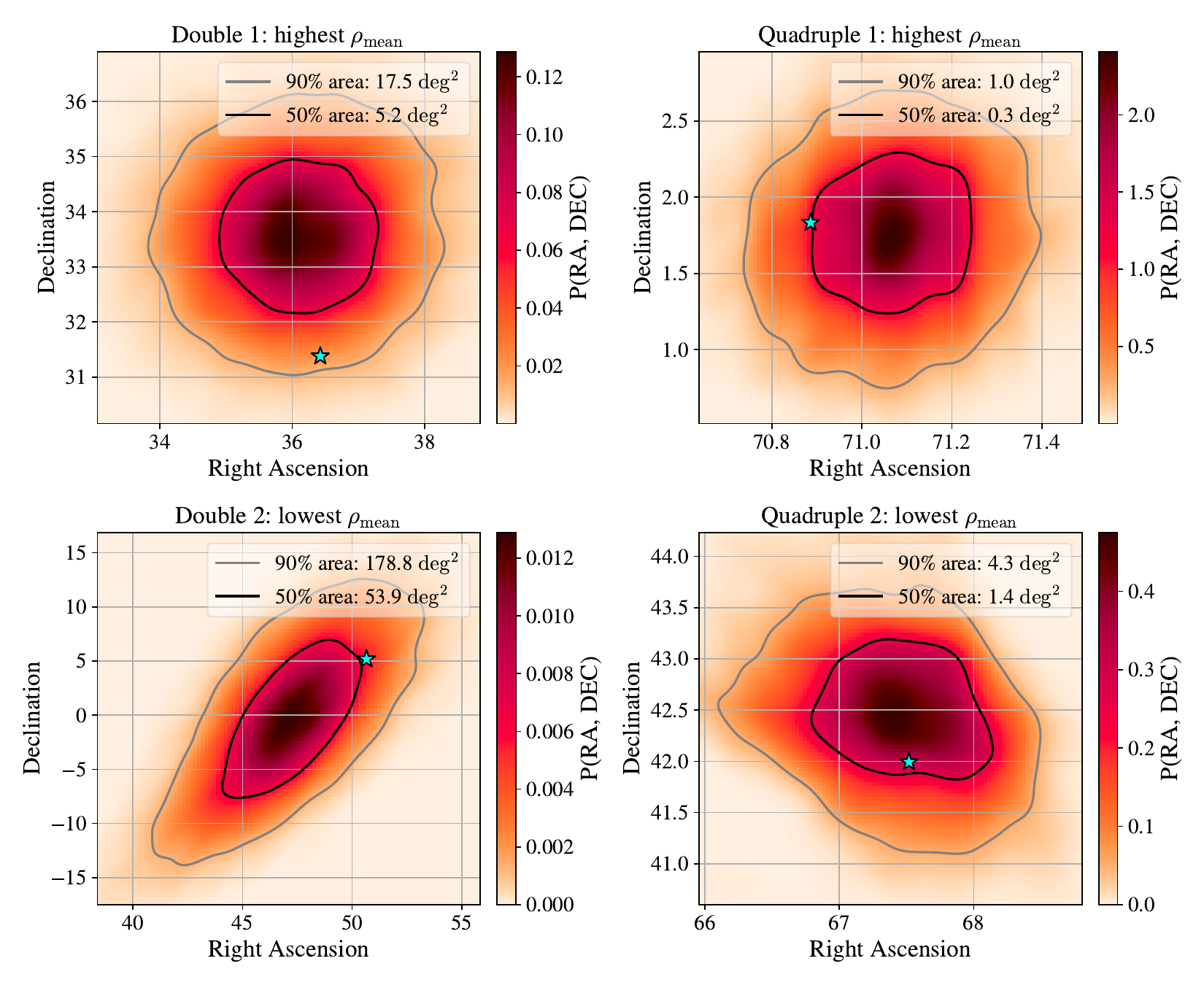}
    \caption{Sky localization maps for selected double- and quadruple-image systems.
    The left (right) panels show double (quadruple) image systems.
    The top row corresponds to systems with the highest $\rho_{\rm mean}$, which exhibit the most tightly constrained sky localizations, while the bottom row shows systems with the lowest $\rho_{\rm mean}$, showing the broader posterior regions.
    The black and gray solid contours represent the 50\% and 90\% credible regions, respectively.
    The star symbols mark the true sky positions of the corresponding GW lensing systems.}
    \label{fig:skyloc}
\end{figure*}

\subsubsection{Method}
If a source-lens system can be uniquely specified, the redshift of the source galaxy ($z_{s}$) can be obtained via EM observations, either through photometric or spectroscopic methods.
The obtained $z_{s}$ can then be used to apply the Hubble-Lema\^{i}tre law ($H_{0} = v_{r} / d_{p}$), where
\begin{equation}
v_{r} (z) = \int^{z}_{0} \frac{c~dz'}{\sqrt{\Omega_{m}(1+z')^{3} + \Omega_{\Lambda}}} 
\end{equation}
is the recessional velocity of the source galaxy, and $d_{p}$ is its proper distance.
In a flat-$\Lambda$CDM universe, where the scale factor at $z=0$ is set to $a_{0}=1$, the proper distance is equivalent to the comoving distance to the source galaxy ($d_{C} = d_{L}/(1+z_{s})$).
In this study, we ignore the peculiar velocity of the source galaxy.
This assumption can introduce biases in $H_{0}$ measurement using bright sirens, such as binary-neutron stars, but it is reasonable for dark-siren analyses, where host galaxies are typically not located in the nearby Universe ($d_{\rm L} < 200 $~Mpc)~\citep{blake2025role,amsellem2025probing}.

Since the redshift of a galaxy cannot be determined perfectly, we assign uncertainties such that the measured redshift follows a Gaussian distribution, where the the standard deviation is $\sigma_{z}$. For the photometric and spectroscopic methods, we adopt $\sigma_{z}= 0.033(1+z_{s})$ and $\sigma_{z}= 0.001(1+z_{s})$, respectively~\citep{bilicki2016wise}.
Using these estimated redshift distributions, the recessional velocity of the source can be determined, and the measured luminosity distance can be converted into the corresponding comoving distance.
Note that the measured luminosity distance from the $j^{\rm th}$ lensed signal is not the true distance to the source, but is rescaled by the lensing magnification factor, such that $d^{\rm eff}_{{\rm{L}},j} = d_{\rm{L}}/\mu_{j}$, where $\mu_{j}$ is the magnification factor of the $j^{\rm th}$ signal.
Recovering the true distance, therefore, requires \emph{delensing} (i.e., removing the effect of magnification), which involves estimating the individual magnification factors $\mu_{j}$ from lens reconstruction.

\begin{figure*}[t]
    \centering
\includegraphics[width=.8\linewidth]{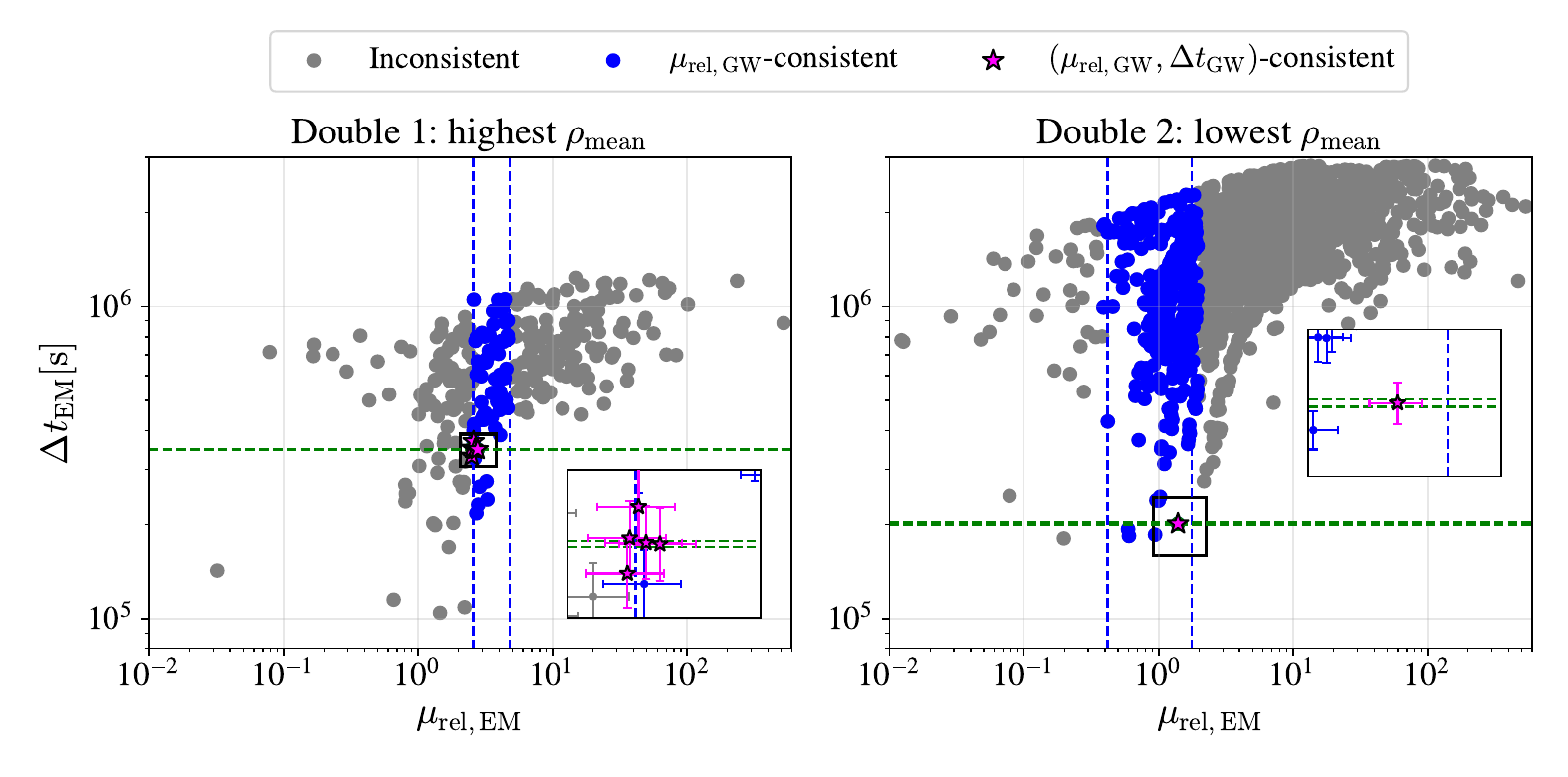}
    \caption{Distributions of the median relative magnification factors and time delays obtained from EM observations for the source–lens galaxy systems within the localized sky area of each double-image case.
    The blue and green dashed lines indicate the constrained boundaries of $\mu_{\rm rel, GW}$ and $\Delta t_{\rm GW}$ inferred from the observed lensed GW signals.
    Blue dots represent lensing systems consistent with the $\mu_{\rm rel, GW}$ constraint, while grey dots denote the systems inconsistent with either constraint.
    Note that no systems are consistent with only the $\Delta t_{\rm GW}$ constraint because the GW time delays are tightly constrained.
    Magenta stars indicate systems consistent with both $\mu_{\rm rel, GW}$ and $\Delta t_{\rm GW}$.
    The inset highlights the error bars on $\mu_{\rm rel, EM}$ and $\Delta t_{\rm EM}$.
    For the lowest (highest) $\rho_{\rm mean}$ cases, five (one) systems remain consistent with the two GW observables.}
    \label{fig:constrained_Nlgw}
\end{figure*}

From the lens-model-independent JPE for the lensed signals~\citep{liu2021identifying,janquart2021fast,janquart2023return,lo2023bayesian}, one can obtain not only source parameters, but also lensing observables for each pair of signals.
For a double-image system, this corresponds to a single triplet, ($\mu_{\rm rel,12}, \Delta t_{12}, \Delta \phi_{n,12}$).
For a quadruple-image system, three such triplets are theoretically available, ($\mu_{\rm rel,1j}, \Delta t_{1j}, \Delta \phi_{n,1j}$) with $j=2,3,4$, although fewer triplets may be obtained if some images are missing or undetectable.

These lensing observables can then be converted into model-dependent lens parameters either analytically~\citep{wright2023determination} or numerically using a rejection-sampling framework~\citep{seo2024inferring}. 
Once the lens parameters are constrained, one can further utilize shape parameters from EM observations, such as the axis ratio ($q$) and elliptical angle ($\phi_{e}$) of the lens galaxies, as well as the inferred redshifted lens mass, to aid in uniquely identifying the source-lens system within the sky localization. This approach relies on the assumption that both the lens galaxy and the lensed source images are included in the catalog. 
Therefore, we use LGC0 for this case. 
When accounting for catalog incompleteness by imposing $m_{\rm th}$, both the lens galaxy and the lensed source galaxy images must be contained within LGC1.

With the lens parameters determined, we can reconstruct the lens galaxy under a given lens model to obtain individual magnification factors, which are used to delens the apparent luminosity distance and retrieve the true distance to the source.
However, it is important to note that the rejection-sampling approach does not always yield accurate lens reconstructions, owing to degeneracies between lensing observables and the underlying lens properties, as well as from the limited availability of EM information due to survey magnitude limits.
This issue will be addressed in Sec.~\ref{subsec:bias}.
\subsubsection{Two scenarios: Double- and Quadruple-image system}
Among the simulated double- and quadruple-image systems, we present four representative detected cases: For each system type, (1) the system with the highest mean S/N ($\rho_{\rm mean}$), defined as the average of the network S/N values of its lensed signals, and (2) the system with the lowest $\rho_{\rm mean}$.

Assuming comparable $\rho_\mathrm{mean}$ for the same detector network, the sky localization of a double-image system improves relative to that of an unlensed event, while a quadruple-image system provides further improvement since more independent measurements of the same event are available.
Figure~\ref{fig:skyloc} shows the skymap posteriors for the four systems.
Considering the complete lensed galaxy catalog (LGC0), the inferred 90\% credible sky areas ($\Delta \Omega_{90\%}$) contain $\mathcal{O}(1)$ and $\mathcal{O}(2-3)$ galaxies with properties compatible with the GW observables for double- and quadruple-image systems, respectively.
The numbers of identified galaxes in each case are summarized in Table~\ref{tab:snr_skydeg_N}.

\begin{table}[t]
\centering
\caption{The mean S/N values ($\rho_{\rm mean}$), 90\% credible sky area ($\Delta \Omega_{90\%}$), numbers of galaxies within these area ($N_{\rm G-G}$), and reduced numbers of galaxies whose properties are consistent with those inferred from the lensed GW signals for each system ($N'_{\rm G-G}$).
As expected, only a few lens candidates are identified within the sky areas constrained by quadruply lensed GW signals.}\label{tab:snr_skydeg_N}
\begin{tabular*}{\linewidth}
{@{\extracolsep{\fill}} lcccc}
\toprule
  \textbf{System} & $\rho_{\rm mean}$ & $\Delta\Omega_{90\%} [\rm{deg}^{2}]$ & $N_{\rm G-G}$ &  $N'_{\rm G-G}$\\
\hline
\textsc{Double 1} & 22 & 17.5 & 306 & 5 \\ 
\textsc{Double 2} & 8.1 & 178.8  & 1971 & 1 \\ 
\textsc{Quadruple 1} & 30.0 & 0.97  & 1 & 1 \\ 
\textsc{Quadruple 2} & 9.0 & 4.35  & 13 &  1 \\ 
\hline
\end{tabular*}
\end{table}

\begin{figure*}[t]
    \centering
\includegraphics[width=0.8\linewidth]{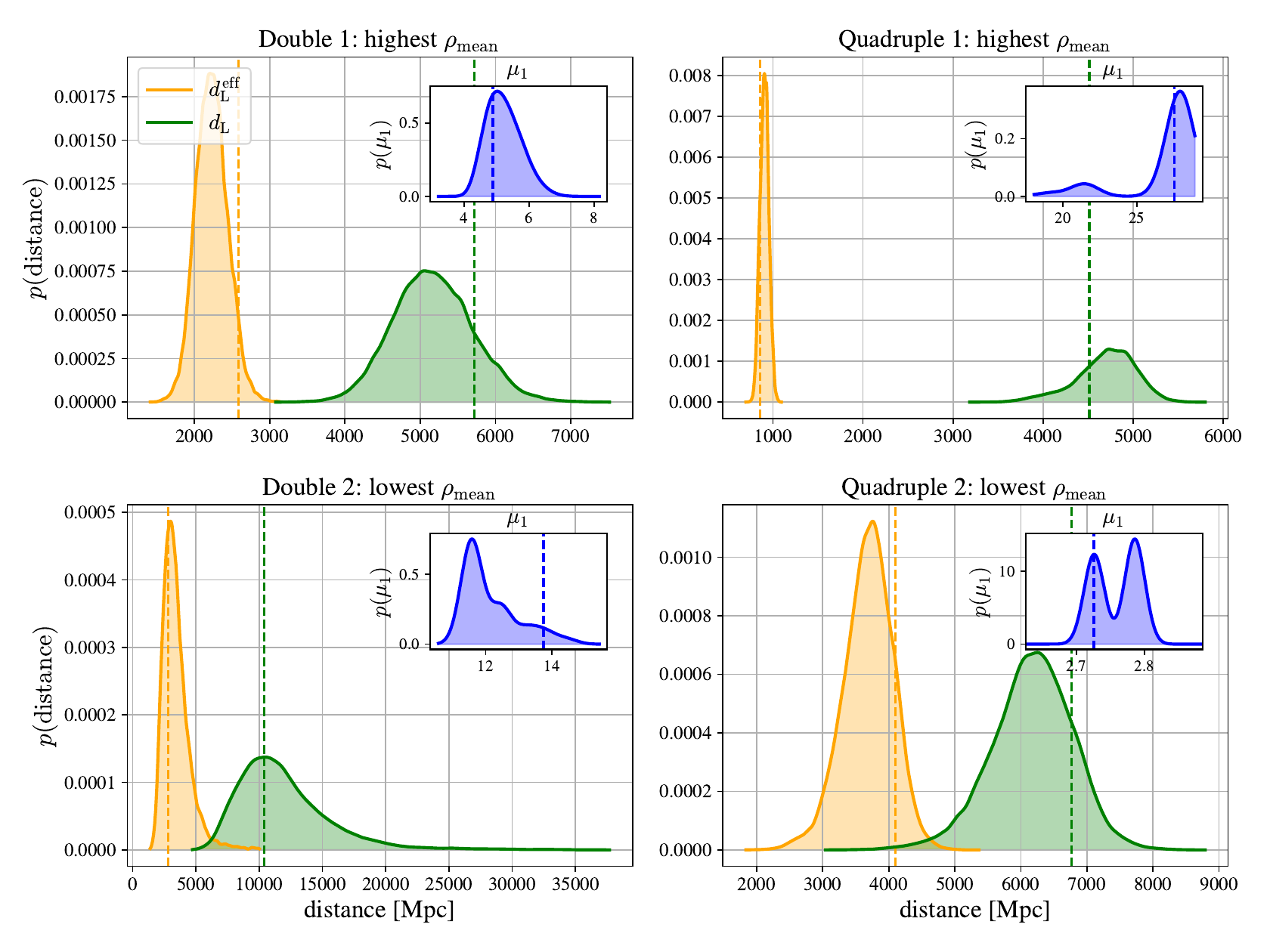}
    \caption{Posteriors of the apparent (orange) and retrieved true (green) luminosity distances for each case, along with the recovered magnification factor (blue), $\mu_{1}$,  for the first-arriving lensed GW signal obtained via rejection sampling.
    Dashed lines indicate the injected values.
    In all cases, the injections are included within the posteriors, and the uncertainties of the retrieved true distances are notably narrower for quadruple-image systems compared to double-image systems.}
    \label{fig:delensing}
\end{figure*}
\begin{figure*}[ht]
    \centering
\includegraphics[width=0.8\linewidth]{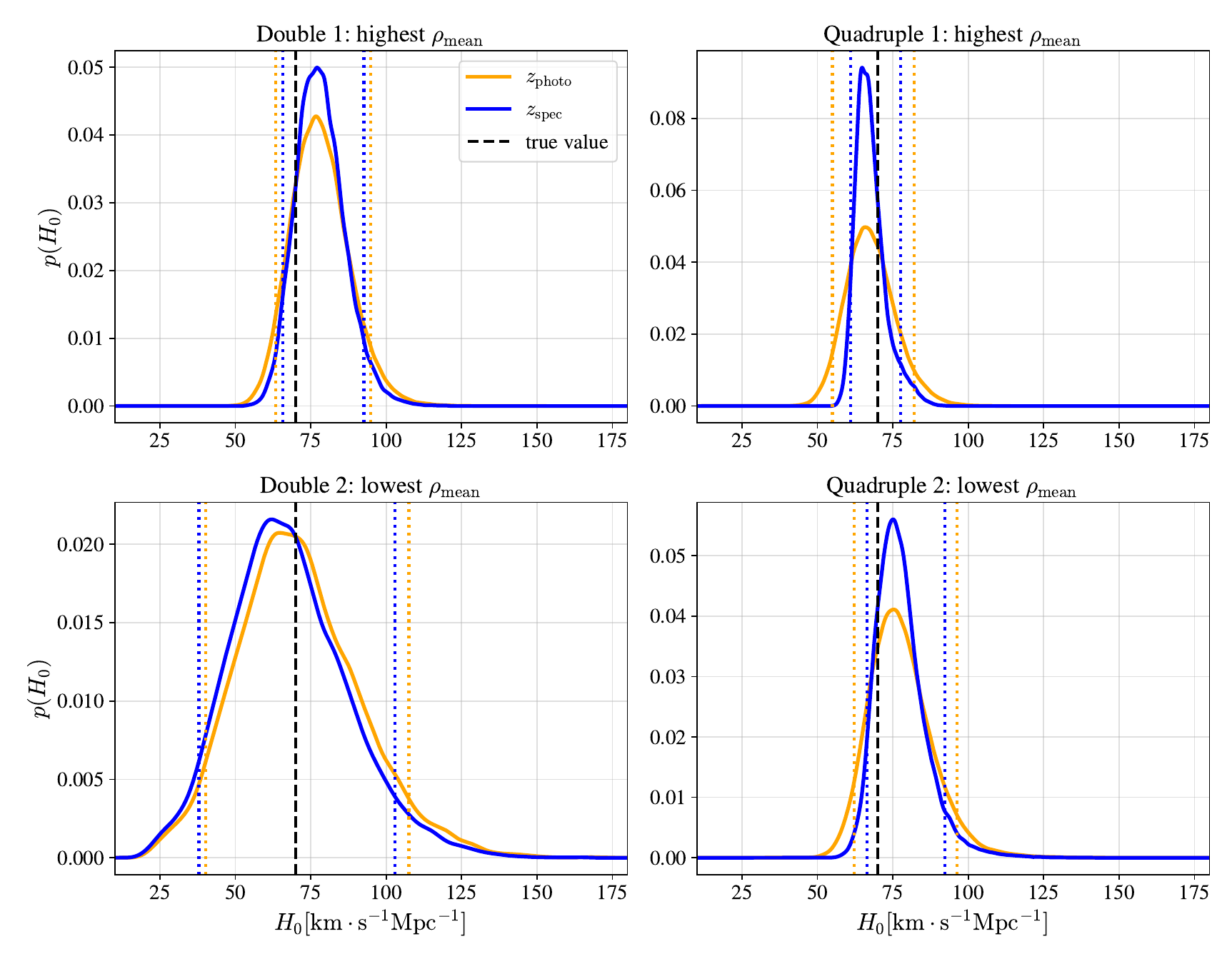}
    \caption{The posteriors of the Hubble constant for Cases 1 and 2 for both double- and quadruple-image systems.
The solid dashed line indicates the true value, and the dotted lines denote the 90\% credible intervals (C.I.).
The blue (orange) solid curve shows the case in which the redshift used to compute the recessional velocity is obtained from spectroscopic (photometric) measurements.
Both Cases 1 and 2 of the quadruple-image system show well-constrained posteriors, as the luminosity distances are precisely recovered and the delensing process performs well.
In contrast, both Cases 1 and 2 of the double-image system exhibit broader C.I. and peaks that deviate further from the true value due to less effective delensing.
In addition, Case 2 of both the double- and quadruple-image systems shows broader C.I. because the precision of distance inference is lower for low-S/N signals.} \label{fig:HL_H0_pos}
\end{figure*}

To further narrow down the lensing system candidates, we utilize the lensing observables, particularly relative magnification factors ($\mu_{\rm rel, EM}$) and time delays.
Since those quantities can be directly obtained from the lensed galaxy catalog, we can compare the EM-derived properties of the source-lens system with those inferred from the lensed GW signals.

As shown in Figure~\ref{fig:constrained_Nlgw}, among the double-image systems in LGC0, the number of systems consistent with the inferred lensing observables reduces to at most a few.
For GW signals, since the arrival time is precisely measured, the time delay between two images provides a stronger constraint than the relative magnification factors inferred from the apparent luminosity distances.
In practice, it is generally challenging to uniquely identify the lens–source pair in double-image configurations in order to measure the source-galaxy redshift, except in cases of high S/N events with correspondingly tight sky localization (e.g., $\sim 20~\mathrm{deg}^2$)~\citep{Hannuksela:2020xor,wempe2024detection}.
In contrast, quadruple-image systems provide much tighter constraints, enabling the source–lens system to be uniquely identified among only a few $N_{\rm G-G}$ candidates as shown in Table~\ref{tab:snr_skydeg_N}.

\begin{figure*}[ht]
    \centering
\includegraphics[width=0.8\linewidth]{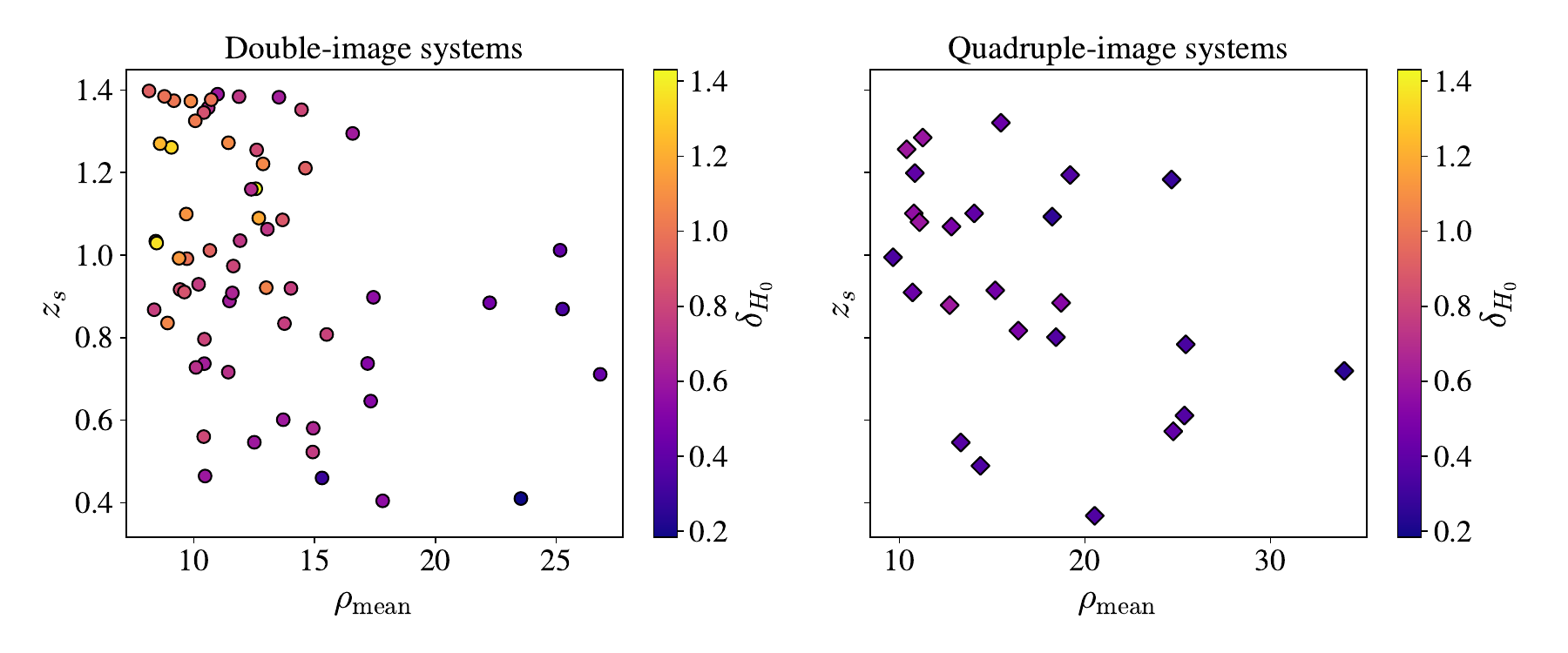}
    \caption{Scatter plots showing the distributions of the mean S/N ($\rho_{\rm mean}$) and source redshift ($z_{s}$) for all detected doubly- (circle markers) and quadruply-lensed (diamond markers) GW systems in our sample pool, color-coded by the relative uncertainty ($\delta_{H_{0}}$) value.
    Generally, the quadruple image systems have lower $\delta_{H_{0}}$ values, reflecting the more informative constraints they provide on $H_{0}$.
    In both cases, events detected with higher $\rho_{\rm mean}$ and located at lower $z_{s}$ tend to exhibit lower $\delta_{H_{0}}$ values, indicating that the corresponding $H_{0}$ posteriors are more tightly constrained.} \label{fig:score_dist}
\end{figure*}
Despite these challenges in the double-image cases, here, we assume that the source–lens pair can be uniquely identified in each system.
Once the lensing systems are uniquely identified, we can further exploit the image positions of the lensed source galaxy and lens galaxy's shape parameters from EM observations to more tightly constrain the other lens parameters, especially $y$, thereby enabling the more accurate computation of individual magnification factors\footnote{In general, the galaxy, as an extended source, does not share the same source-plane position $y$ as the GW event embedded within it, i.e., $y_{\rm gal} \neq y_{\rm gw}$. Since the magnification factor is highly sensitive to $y$, it is necessary to adopt $y_{\rm gw}$ rather than $y_{\rm gal}$.}.
Following \citet{seo2024inferring}, we assume that the shape parameters ($q, \phi_{e}$) can be determined with an uncertainty of $\pm 5\%$ and that the image positions can be measured with an uncertainty of $0.1''$ from EM observations.

The inferred individual magnification factors are shown in Figure~\ref{fig:delensing}.
These distributions are sampled to retrieve the true luminosity distances from the previously obtained apparent distances following Eq.~\eqref{eq:lensed_observables}.
Specifically, we draw random values from the posterior distribution of the magnification factor and, for each draw, multiply it by a randomly sampled value from the apparent-distance posterior to obtain a corresponding realization of the true luminosity distance.
As expected, lens reconstruction (i.e., the inference of individual magnification factors) is less accurate for double-image systems, for which only two constraints ($\mu_{\rm rel}$ and $\Delta t$) are available, than for quadruple-image systems, which provide four additional constraints.

Using the recovered true luminosity distances and the photometric or spectroscopic redshifts of the source galaxies in each identified lensing system, we compute the corresponding comoving distances and recessional velocities. 
The Hubble–Lemaître law is then applied to derive $H_{0}$.
Figure~\ref{fig:HL_H0_pos} shows $H_{0}$ posteriors obtained from the four cases.
The 90\% credible intervals (C.I.) of the posteriors obtained from quadruple-image systems with spectroscopic redshifts are much narrower than in the other cases.

To quantify how well the posteriors ($p(H_{0})$) are converged, we define a relative uncertainty ($\delta_{H_{0}}$), as
\begin{equation}
    \delta_{H_{0}} \equiv \frac{90\% \; \mathrm{C.I.}}{H_{0,\rm med}}, \label{eq:score}
\end{equation}
where $H^{\rm med}_{0}$ denotes the median of the posterior.
Normalizing the credible interval by the median, rather than the mean, which is more sensitive to skewed or long-tailed posteriors, provides a dimensionless measure of the fractional uncertainty, analogous to a coefficient of variation.
A lower value of $\delta_{H_{0}}$, therefore, indicates a more tightly constrained posterior.

For each system, the estimated Hubble constants with their 90\% C.I. and the corresponding $\delta_{H_{0}}$ values are listed in Table~\ref{tab:score_CI}.
Among systems with the same number of images, a higher $\rho_{\rm mean}$ results in tighter $H_{0}$ posteriors and consequently a lower $\delta_{H_{0}}$ value, because high-S/N events produce narrower posteriors for the recovered luminosity distance.
For comparable S/N, quadruple-image systems yield narrower posteriors than double-image systems, leading to systematically lower $\delta_{H_{0}}$ values, as four lensed signals provide more distance realizations that improve the precision of the recovered luminosity distance.

\begin{table}[t]
\caption{Relative uncertainties and the corresponding Hubble constant posteriors, along with their 90\% C.I., inferred using spectroscopic redshifts for the double- and quadruple-image lensing systems. }\label{tab:score_CI}
\centering
\begin{tabular*}{\linewidth}
{@{\extracolsep{\fill}} lcc}
\toprule
  \textbf{System} & $H_{0}$ $[\rm{km}\cdot s^{-1}\cdot Mpc^{-1}]$&   $\delta_{H_{0}}$  \\
\hline
\textsc{Double 1} & $77.7 ^{+14.6}_{-11.8}$ & 0.34    \\ 
\textsc{Double 2} & $67.4^{+48.4}  _{-33.6}$   &  1.22\\ 
\textsc{Quadruple 1} & $66.3 ^{+7.2}  _{-5.7}$ & 0.19     \\ 
\textsc{Quadruple 2} & $76.5 ^{+16.3}_{-10.0}$ & 0.34  \\ 
\hline
\end{tabular*}
\end{table}

The low S/N of the lensed GW signals indeed contributes significantly to the large uncertainties, as the inferred apparent luminosity distances are often poorly constrained.
However, for sources located at high redshift, the uncertainty can be large regardless of the S/N of the detected lensed signals.
Such high-$z$ detections generally fall into two representative cases: (1) light BBH events that become observable only after undergoing strong magnification, and (2) intrinsically heavy BBH events that require only moderate magnification to exceed the detection threshold.
In the first case, the apparent luminosity distances are initially well constrained; however, once delensing is applied to recover the true luminosities, the variance of the recovered values increases in proportion to the magnification factor. 
In the second case, the apparent distances are already broadly distributed.
In both cases, the resulting $H_{0}$ posteriors are expected to be poorly constrained.

Figure~\ref{fig:score_dist} shows how the $\delta_{H_{0}}$ varies across the $\rho_{\rm mean}$ and GW-source redshift distributions of all simulated, detected lensed signals.
For both doubly- and quadruply-lensed systems, there is a tendency for systems with lower $\rho_{\rm mean}$ and higher-$z_{s}$ to exhibit larger $\delta_{H_{0}}$ values.
At high redshift, the double-image systems generally show uniformly higher $\delta_{H_{0}}$ values, whereas for the quadruple-image systems, the uncertainty is dominated by low $\rho_{\rm mean}$ rather than high $z_{s}$, which indicates that detecting quadruple-image systems is more beneficial for constraining $H_{0}$ from distant sources.

Beyond statistical uncertainties regarding the detector's sensitivity, inaccurate delensing can also introduce additional errors, even when the apparent luminosity distance is well-constrained, and the source lies in a relatively low-$z$ universe.
Such errors may occur during rejection sampling due to degeneracies between lensing observables and lens parameters.
If the individual magnification factors obtained from delensing, which are used to recover the true luminosity distance, are not tightly constrained, the resulting distance posterior will broaden rather than remain sharply peaked around the true value.
Hence, it is important to constrain the delensed luminosity distances as tightly as possible, to ensure precise $H_0$ inference.

In this section, we assume that all galaxy-galaxy lensing events are included in the catalog (i.e., complete catalog) and that each lensing system can be uniquely identified, implying that delensing is expected to be highly accurate and the associated uncertainty can be neglected.
In practice, however, some lensing systems may have multiple candidate host galaxies within the sky-localization region, or some hosts may be too faint to be in the catalog, introducing additional uncertainty in the inferred $H_{0}$ for individual events. 
When multiple lensed dark sirens from different lensing systems are detected, the situation becomes more complex.
In the following section, we describe how to combine information from multiple dark sirens while accounting for catalog incompleteness, the redshift distributions of potential host galaxies, and imperfect delensing.

\subsection{Multiple lensed dark sirens and galaxy catalog}\label{subsec:gwcosmo}
When multiple lensed dark sirens originating from different lensing systems are detected, the individual $H_{0}$ posteriors discussed in Sec.~\ref{subsec:HLlaw} cannot be simply combined.
Instead, a joint likelihood must be constructed and multiplied by the $H_{0}$ prior to obtain the full posterior distribution.
In this context, we adapt \texttt{gwcosmo}~\citep{PhysRevD.101.122001,gray2022pixelated,gray2023joint}\footnote{The \texttt{gwcosmo} code is available at \url{https://git.ligo.org/lscsoft/gwcosmo}.}, a pipeline designed for cosmological inference and population studies using GW events, to estimate the $H_{0}$ posterior for lensed GW observations.
\subsubsection{Method}\label{subsubsec:gwcosmo_method}
Given that we detect $N_{\rm lens, det}$ lensed GW systems, each consisting of multiple lensed signals denoted collectively as $\{x_{\rm lgw}\}$ with corresponding detection flag $\{D_{\rm lgw}\}$, the  posterior of $H_{0}$ can be written as
\begin{equation}
\begin{aligned}
\label{pos_h0_simple}
p(H_{0} \mid \{x_{\rm lgw}\}, \{D_{\rm lgw}\})&\propto  
p(H_{0}) \, p(N_{\rm lens, det} \mid H_{0}) \\
&\times \prod_{i=1}^{N_{\rm lens, det}} \mathcal{L}(x_{{\rm lgw},i} \mid  D_{{\rm lgw},i}, H_{0}),
\end{aligned}
\end{equation}
where $p(H_{0})$ is the prior of the Hubble constant and $p(N_{\rm lens, det}|H_{0})$ is the probability of detecting $N_{\rm{lens,det}}$ GW lensing systems for a given value of the Hubble constant.
In the unlensed case, the detection count term $p(N_{\rm det} | H_{0})$ can be treated as independent of $H_{0}$ by setting the prior on the binary merger rate to be $p(R) \propto 1/R$.~\citep{fishbach2018does,PhysRevD.101.122001,gray2023joint}.
However, when lensing is considered, the detection probability becomes a function of magnification and lensing optical depth, both of which depend on $H_{0}$.
Hence, $p(N_{\rm lens,det} | H_{0})$ retains a dependence on $H_{0}$ through the lensing rate.
The last term in Eq.~\eqref{pos_h0_simple} is the joint likelihood functions of the $N_{\rm lens,det}$ events, where a single likelihood can be written as
\begin{equation}
\label{eq:individual_likeli}
 \mathcal{L}(x_{{\rm lgw}} \mid D_{{\rm lgw}}, H_{0}) = \frac{p(x_{{\rm lgw}} \mid H_{0})}{p(D_{{\rm lgw}} \mid H_{0})},
\end{equation}
where 
\begin{equation}
    p(D_{{\rm lgw}} \mid H_{0}) = \int p(D_{{\rm lgw}} \mid x_{{\rm lgw}},H_{0}) 
    \times p(x_{{\rm lgw}} \mid H_{0}) {\rm{d}}x_{{\rm lgw}}.
\label{eq:Dlgw}
\end{equation}
The term $p(D_{{\rm lgw}} \mid x_{{\rm lgw}}, H_{0})$ takes the value of 1 if $x_{\rm lgw}$ is identified as a GW lensing system, and 0 otherwise.
A system is identified as strongly lensed when multiple GW signals are detected with $\rho_{\rm net} > 8$ and are mutually consistent in their intrinsic source parameters and sky localization, with a sufficiently large Bayes factor favoring the strong-lensing hypothesis.
In practice, however, the finite duty cycle of GW detectors implies that one or more signals in a lensed system may be missed, resulting in cases where only a single image is detected if the detectors are not operating at the corresponding arrival times.
In such cases, the detected signal may not be recognized as lensed and is therefore treated as an unlensed event, which can introduce a bias in the inferred value of $H_0$. 
We discuss this effect in Sec.~\ref{sec:challenges} in more detail.

The term $p(x_{\rm lgw} \mid H_{0})$ represents the probability of obtaining the lensed GW data for a given value of the $H_{0}$.
In practice, $p(x_{\rm lgw} \mid H_{0})$ depends on all parameters that determine the S/N of GW event, which includes not only $H_{0}$, but also lens and GW source parameters ($\theta_{l}, \theta_{s}$)\footnote{$\theta_{s}$ denotes the intrinsic source parameters, excluding the redshift $z_{s}$. Likewise, $\theta_{l}$ refers to the lens parameters other than $z_{l}$.}, redshifts ($z_{s},z_{l}$), and sky patches ($\Omega$) constituting the sky localization map.
Hereafter, all dependencies on $x_{\rm lgw}$ are written explicitly.
With this convention, Eq.~\eqref{eq:individual_likeli} can be rewritten in the following form
\begin{widetext}
\begin{equation}
\label{eq:single_L_full}
\begin{aligned}
    \mathcal{L}(x_{{\rm lgw}} | D_{{\rm lgw}}, H_{0}) &=  \left[ \iiiint \sum^{N_{\rm pix}}_{j} p(D_{{\rm lgw}} | \Omega_{j}, \theta_{s}, \theta_{l}, z_{s}, z_{l}, H_{0})p(\theta_{s}, \theta_{l}, z_{s}, z_{l},|\Omega_{j},H_{0})p(\Omega_{j})d\theta_s d\theta_{l} dz_{s} dz_{l} \right]^{-1}\\
    & \times \iiiint \sum^{N_{\rm pix}}_{j} p(x_{{\rm lgw}} | \Omega_{j}, \theta_{s}, \theta_{l}, z_{s}, z_{l}, H_{0}) p( \theta_{s}, \theta_{l}, z_{s}, z_{l},|\Omega_{j},H_{0})p(\Omega_{j}) d\theta_s d\theta_{l} dz_{s} dz_{l},
\end{aligned}
\end{equation}
\end{widetext}
where the term $p(x_{\rm lgw}|...)$ is represented in a pixelated form by partitioning the sky-localization map into discrete patches on a \textsc{HEALPix} map~\citep{Healpix,healpy}\footnote{\url{http://healpix.sf.net}.}, such that each sky patch is assigned the subset of likelihood samples whose inferred sky locations fall within that region~\citep{gray2022pixelated}.
The pixelation scheme is adopted to account for sky-varying incompleteness in galaxy catalogs.
Under this discretization, the integral over sky position is replaced by a sum over pixels.

The term $p(\theta_{s}, \theta_{l}, z_{s}, z_{l},|\Omega_{j}, H_{0})$ denotes the joint prior distribution of the source and lens parameters and the redshifts of the source and lens for a given $H_{0}$.
In contrast to the analysis of single dark-siren events described in Sec.~\ref{subsec:HLlaw}, the redshifts of the source and lens galaxies are inferred statistically by cross-matching the sky localization of the detected lensed GW signals with a galaxy-galaxy lensing catalog, such as LGC0, rather than being obtained through a unique host identification.
For a given set of sky patches, the redshift distributions of the source and lens galaxies are obtained by selecting all catalog galaxies that lie within each patch. 
These patch-specific redshift distributions define the \emph{line-of-sight (LoS) redshift prior}, which is independent of the intrinsic distributions of the GW source parameters.
The joint prior can therefore be written as
\begin{equation}
\label{eq15}
    p(\theta_{s}, \theta_{l}, z_{s}, z_{l},|\Omega_{j}, H_{0}) = p(\theta_{s}| H_{0}) p(z_{s},z_{l},\theta_{l}|\Omega_{j}, H_{0}).
\end{equation}
The intrinsic properties of the GW source (e.g., masses, spins, and orbital parameters) are independent of sky location, and the first term carries no explicit dependence on $\Omega_{j}$.

The denominator in Eq.~\eqref{eq:single_L_full} accounts for selection effects arising from the finite sensitivity of GW detectors, which makes louder signals more likely to be detected.
Accurate modeling of these effects is required to avoid selection biases (e.g., Malmquist bias) and to obtain unbiased inferences of population-level parameters~\citep{farr2019accuracy}.
The lensed GW selection effect can be characterized as the fraction of all GW sources that, after being lensed by intervening objects, are detected with network S/N exceeding the detection threshold, which we set to $\rho_{\rm thr}= 8$.
For doubly lensed systems, both lensed GW signals must individually exceed $\rho_{\rm thr}$, whereas for quadruply lensed systems, at least two images must exceed this threshold, as a minimum of two detections is required to confidently identify a strongly lensed event.

In principle, one may perform a dedicated sub-threshold search to identify a counterpart of a detected super-threshold GW signal that is insufficiently magnified or strongly demagnified to be detected above the threshold~\citep{li2023targeted,li2025tesla}.
Furthermore, a lensed GW signal could be identified even from a single image by searching for waveform distortions in Type-II images induced by higher-order GW modes~\citep{ezquiaga2021phase,janquart2021identification}. 
However, such effects are not directly relevant to the high-magnification regime or to improvements in sky localization, and therefore, we do not consider this possibility in this work.

In terms of EM selection effects, although a complete catalog is assumed in Sec.~\ref{subsec:HLlaw}, realistic galaxy surveys may fail to include lensed source-galaxy images or lens galaxies whose apparent magnitudes fall below the survey’s detection threshold.
The resulting incompleteness of optical surveys introduces additional selection effects, as part of the redshift information entering the LoS redshift prior is missing.
To account for the selection effects introduced by an incomplete galaxy catalog, we construct the LoS redshift prior by marginalizing over the absolute and apparent magnitudes of both the source and lens galaxies, based on the method in \cite{gray2023joint}.
Since the LoS prior is defined for galaxies that are potential hosts of the GW source within the sky localization region, we introduce binary parameters $s$ and $l$, denoting whether a given source galaxy hosts a GW source and whether detected lensed GW signals arise from the same lensing system, i.e., whether they are genuine counterparts corresponding to multiple lensed images of a single underlying GW source.

As described in Sec.~\ref{subsec:gc}, we impose an apparent-magnitude threshold $m_{\rm th}$ to construct incomplete galaxy catalogs from UGC0 and LGC0.
We choose $m_{\rm th} = 21.87$ such that UGC1 contains approximately half of the galaxies in UGC0, corresponding to \(\sim 3.1\times10^{7}\) galaxies.
Under the same magnitude cut, LGC1 contains about 34\% of the galaxy-galaxy lensing systems in LGC0.
Using the four simulated galaxy catalogs, we compute the LoS redshift priors on a HEALPix grid with \(\mathrm{n_{side}}=128\) for UGCs, chosen to ensure consistency with \citet{gray2023joint}, and \(\mathrm{n_{side}}=64\) for LGCs, as the number of objects drastically decreases due to the small strong-lensing optical depth.

By substituting the joint likelihood of the lensed GW signals from Eq.~\eqref{eq:single_L_full} into Eq.~\eqref{pos_h0_simple}, and adopting the expressions in Eqs.~\eqref{eq15}, the posterior for $H_{0}$ can be written explicitly as
\begin{widetext}
\begin{equation}\label{main_h0_pos}
\begin{aligned}
&p(H_{0} | \{x_{\rm lgw}\}, \{D_{\rm lgw}\}) \propto p(H_{0}) p(N_{\rm lens, det} | H_{0})\\
&\times \left[ \iiiint  p(D_{\rm lgw} |  \theta_{s}, \theta_l, z_{s}, z_{l}, H_{0}) p(\theta_s | H_{0}) \sum_{j}^{N_{\rm pix}} p(z_{s},z_{l}, \theta_l | \Omega_j, H_{0},s,l) d\theta_l d\theta_{s} dz_{l} dz_{s} \right]^{-N_{\rm lens,det}}\\
&\times \prod_{i}^{N_{\rm lens, det}} \left[\iiiint  \sum_{j}^{N_{\rm pix}} p(x_{{\rm lgw}, i} | \Omega_j, \theta_{s},\theta_l, z_{s}, z_{l} ,H_{0}) p(\theta_{s} | H_{0}) p(z_{s},z_{l}, \theta_l | \Omega_j, H_{0},s,l) d\theta_l d\theta_{s} dz_{l}dz_{s} \right]
\end{aligned}
\end{equation}
\end{widetext}
A detailed derivation of Eq.~\eqref{main_h0_pos}, together with an explanation of the contribution of each term, is presented in Appendix~\ref{app:H0posterior}.

\subsubsection{Impact of strong lensing on $H_{0}$ measurement}

\begin{figure*}[ht]
    \centering
\includegraphics[width=0.8\linewidth]{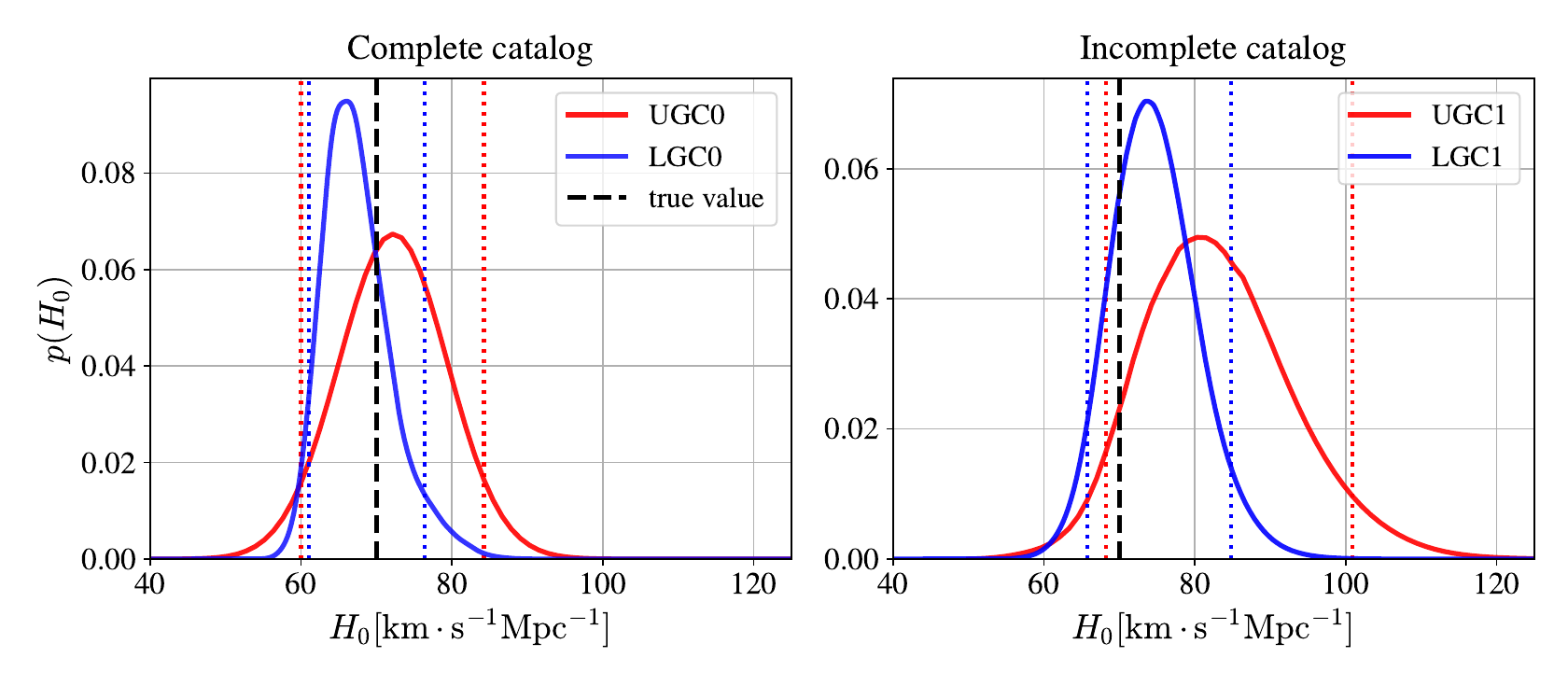}
    \caption{Posteriors of $H_{0}$ for the four analysis scenarios.
    The left panels use complete galaxy catalogs (UGC0 and LGC0), while the right panels incorporate catalog incompleteness (UGC1 and LGC1).
    Unlensed and lensed cases are shown by solid red and solid blue lines, respectively.
    The dashed black lines indciates the true value of $H_{0}$.
    In all cases, spectroscopic redshift measurements are employed.
    Although the lensed cases comprise only 8 events compared with 250 unlensed signals, they still produce tighter posteriors. Incomplete catalogs broaden the posteriors due to increased uncertainty in host-galaxy identification, while less precise delensing in the lensed cases increases the luminosity-distance uncertainty, further widening the $H_{0}$ posterior.}
    \label{fig:final H0}
\end{figure*}
To estimate $H_{0}$ using the set of unlensed signals together with UGCs and the set of lensed signals together with LGCs, we draw random sub-samples from the simulated population of detectable events to emulate a realistic observing run.
The sample pool comprises 250 unlensed signals, consistent with the number of events detected during the LVK fourth observing run. 
In addition, under an optimistic scenario, we include 6 doubly-lensed and 2 quadruply-lensed systems identified among the same population when 250 unlensed events are detected.
Although this number of lensed events is higher than conservative expectations\footnote{In a realistic observing scenario, a fraction of lensed GW signals can be missed due to the finite detector’s duty cycle and the limited efficiency of search pipelines.}, it is adopted to ensure sufficient statistics for reconstructing smooth posterior distributions while maintaining the expected ratio between double- and quadruple-image systems.

Figure~\ref{fig:final H0} shows the posteriors of $H_{0}$ for four analysis scenarios: unlensed GW signals and lensed GW signals with complete or incomplete galaxy catalogs.
In all cases, the posteriors converge well to the true value, but the 90\% credible interval is noticeably narrower for the lensed signals with complete catalogs.
To be specific, among the lensed signals, doubly-lensed GW signals exhibit broad sky localization (as shown in Figure~\ref{fig:skyloc}), which results in many potential host galaxies along the line of sight and complicates identification of the true lensing system.
In addition, the uncertainty in the luminosity distance remains relatively large, yielding only modest constraints on $H_{0}$.
In contrast, although quadruply-lensed signals are less common, they provide substantially improved sky localization, which significantly reduces the number of candidate host galaxies and the luminosity-distance uncertainty, yielding considerably tighter constraints on $H_{0}$.

For incomplete catalog cases (UGC1 and LGC1), the posteriors are broader than in the corresponding complete-catalog cases, as the uncertainty in identifying the true host galaxy increases.
In particular, delensing accuracy in the LGC1 scenario is further limited by the incomplete information on the lensing systems, which propagates into larger luminosity-distance uncertainty.

Overall, a few strongly lensed dark siren systems analyzed with a dedicated galaxy–galaxy lensing catalog can improve the precision of $H_{0}$ by roughly 50\% compared to the unlensed analysis based on a few hundred dark sirens combined with a galaxy catalog (see Table~\ref{tab:H0_deltaH0} for details). 
To reach a few-percent fractional uncertainty ($\delta_{H_{0}} < 0.1$), comparable to constraints obtained from Type Ia supernova measurements, our lensing analysis framework indicates that several tens of lensed GW systems are necessary under the assumption of a complete catalog. 
For instance, using a randomly selected sample of 32 doubly lensed and 8 quadruply lensed events, we obtain $\delta_{H_{0}} = 0.09$.

\begin{table}[t]
\centering
\caption{Median values and 90\% credible intervals of recovered $H_{0}$ posteriors and their relative C.I. for each catalog case.}\label{tab:H0_deltaH0}
\begin{tabular*}{\linewidth}
{@{\extracolsep{\fill}} lccc}
\toprule
\textbf{Catalog}  & Completeness & \bf{$H_{0}[\rm {km\cdot s^{-1} \cdot Mpc^{-1}}]$} & $\delta_{H_{0}}$\\
\hline
UGC0  & 100  & $72.12_{-12.12} ^{+12.12}$   & 0.34 \\
UGC1     & 50       & $82.73_{-14.55}^{+18.18}$   & 0.40\\
LGC0     &  100    & $66.85^{+9.49 }_{-5.77}$  & 0.23\\
LGC1     &  34    & $74.29^{+10.45}_{-8.53}$ & 0.26\\
\hline
\end{tabular*}
\end{table}
\section{Challenges in measurement due to lensing bias}\label{sec:challenges}

\begin{figure*}[t]
    \centering
    \subfigure[No delensing applied]
    {
        \includegraphics[width=0.8\linewidth]{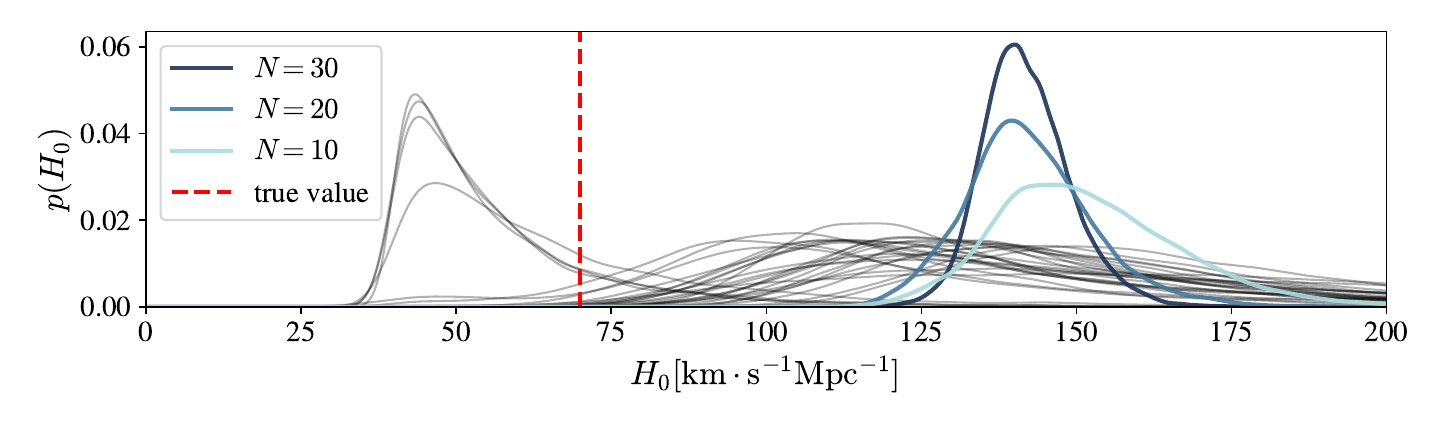}
        \label{no_delens}
    }
    \subfigure[Delensing with SIS model]
    {
        \includegraphics[width=0.8\linewidth]{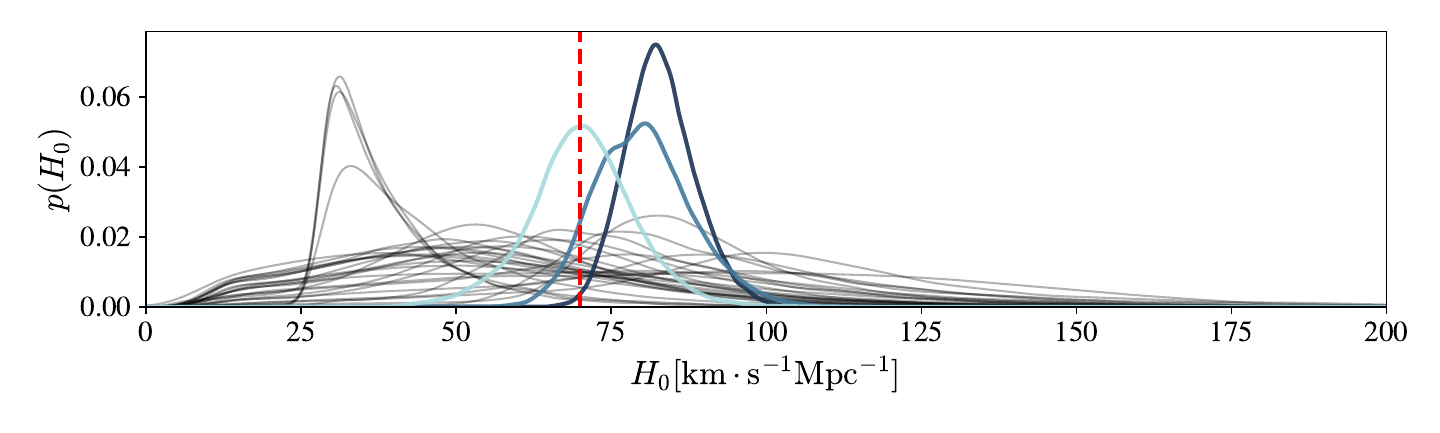}
        \label{SIS_delens}
    }
    \subfigure[Delensing with SIE model]
    {
        \includegraphics[width=0.8\linewidth]{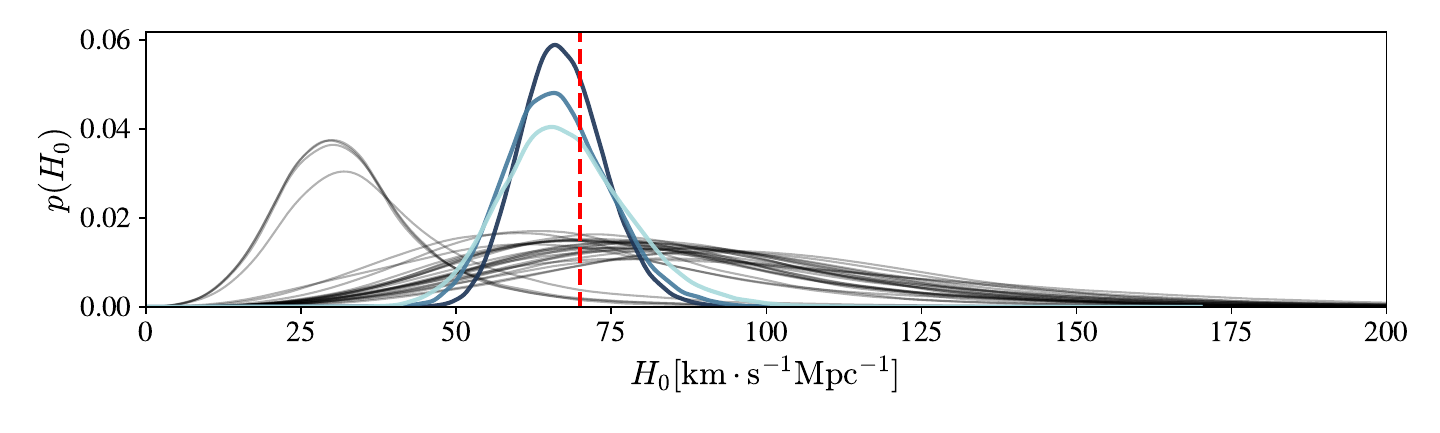}
        \label{SIE_delens}
    }
    \caption{$H_{0}$ posteriors obtained from multiple realizations of a double-image BBH lensing system with a true SIE mass profile.
    Gray curves show the posterior from each individual realization, while the sky-blue ($N=10$), blue ($N=20$), and navy ($N=30$) curves represent the progressively combined posteriors.
    The dashed red lines indicate the true value of $H_{0}$.
    The top, middle, and bottom panels correspond to the cases of no delensing, SIS delensing, and delensing with the correct SIE model, respectively.
    In the absence of delensing, the inferred $H_{0}$ is strongly biased for all realization numbers.
    When delensing is performed with an incorrect SIS model, the bias is less apparent for a small number of realizations but becomes increasingly significant as realizations accumulate, even though the credible intervals shrink.
    In contrast, delensing with the correct SIE model recovers an unbiased estimate of $H_{0}$.}
    \label{fig:bias}
\end{figure*}
\begin{figure*}[t]
    \centering
    \subfigure[$H_0$ inferred from an unlensed population containing hidden lensed signals]
    {
        \includegraphics[width=0.4\linewidth]{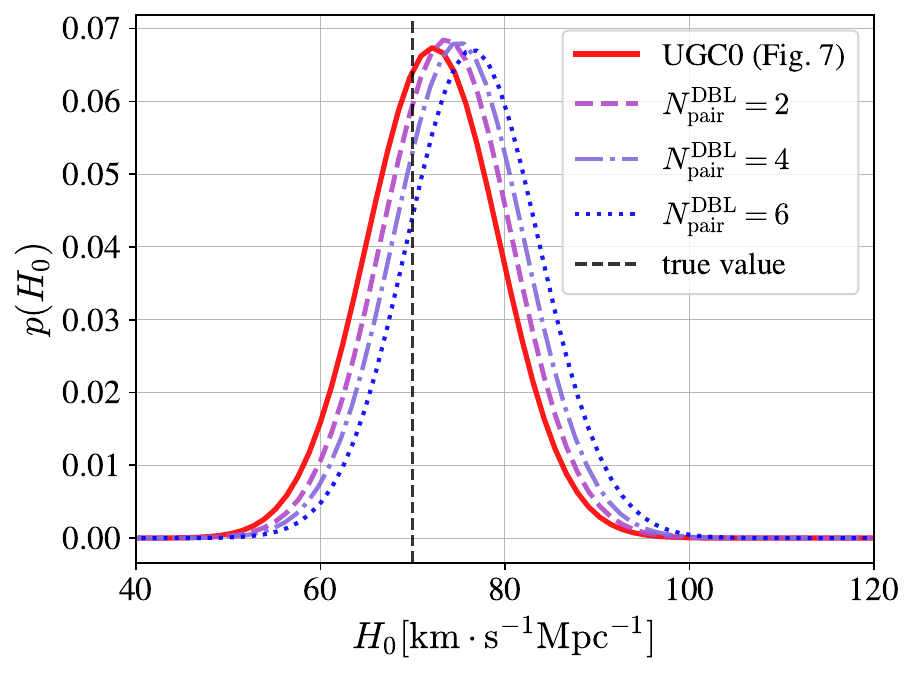}
        \label{L_in_UL}
    }
    \subfigure[$H_0$ inferred from a lensed population containing hidden unlensed signals]
    {
        \includegraphics[width=0.4\linewidth]{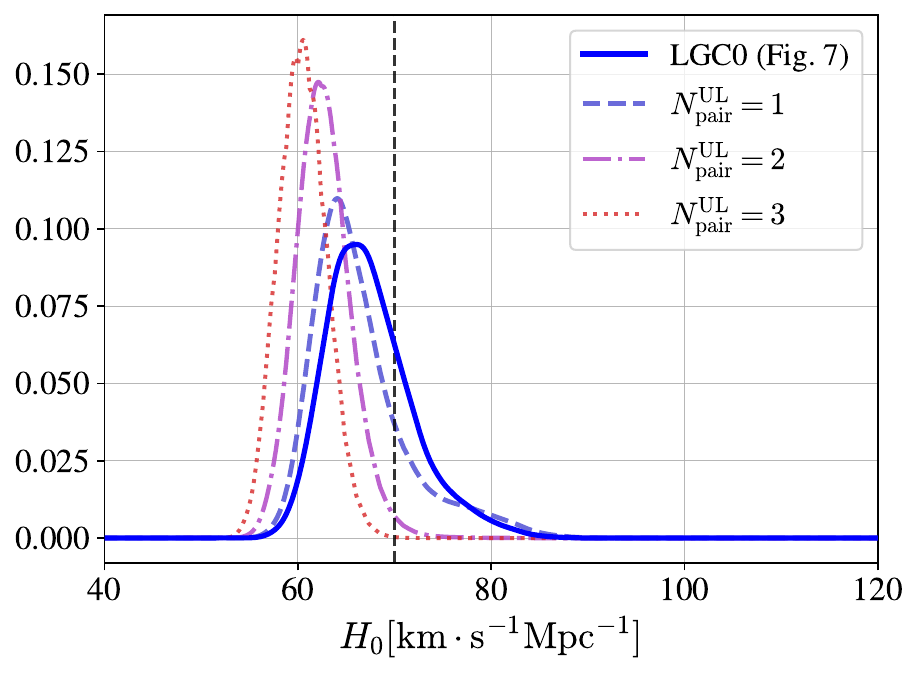}
        \label{UL_in_L}
    }
    \caption{Biased $H_{0}$ posteriors obtained from a subset of misclassified signals within a population where the majority are correctly classified.
    The left and right panels show $H_0$ posteriors biased due to misclassification of unlensed signals as lensed, and lensed signals as unlensed, respectively, with the correctly inferred $H_0$ from Figure~\ref{fig:final H0} shown for reference.
    The dashed black lines denote the true value of $H_{0}$.
    As the number of misclassified events increases, $N^{\rm DBL}_{\rm pair}$ in the left panel and $N^{\rm UL}_{\rm pair}$ in the right panel, the posteriors gradually shift away from the correctly inferred $H_0$.
    In the left panel, they shift from the red solid line to the blue dotted line, while in the right panel, they shift from the blue solid line to the red dotted line.}
    \label{fig:hidden_bias}
\end{figure*}

In this section, we investigate the biases in $H_{0}$ inference when lensing is ignored, i.e., using the conventional unlensed framework, and when the lensing effect is incorrectly interpreted.
\subsection{Systematic bias in lens reconstruction}~\label{subsec:bias}
Accurate measurements of $H_{0}$ using lensed GW signals require a reliable lens reconstruction.
If lensing is ignored or the reconstructed lens fails to fully capture the lensing potential of the foreground galaxy,  for example, due to unresolved substructure, it can introduce significant bias in the inferred parameters, most notably the luminosity distance to the GW source and, consequently, the inferred value of $H_{0}$.
Hence, minimizing such potential systematic biases is essential.

To demonstrate that an incorrect lens reconstruction and delensing procedure introduces a significant bias in the estimation of $H_{0}$, we consider multiple realizations of a single GW lensing system.
Specifically, we adopt the same galaxy-galaxy lensing configuration as the `Double 1' system shown in Table~\ref{tab:snr_skydeg_N}, fixing the lens parameters and distance to the source, while varying the BBH component masses.
This approach allows us to generate multiple independent realizations of the same underlying lensing system.

We simulate 30 such realizations and analyze them under three different assumptions: an unlensed configuration (i.e., no delensing applied), a singular isothermal sphere (SIS) lens model, which represents an incorrect lens model, and an SIE model, corresponding to the correct lens model.
Figure~\ref{fig:bias} shows the posteriors of $H_{0}$ inferred from individual lensed BBH events, together with the progressively combined posteriors under each modeling assumption.
Although potential biases are difficult to diagnose from the posterior of a single realization, combining multiple realizations reveals clear systematic trends.
In the no-delensing case shown in Figure~\ref{no_delens}, the inferred $H_{0}$ is significantly biased for both small and large numbers of combined realizations. 
For the SIE case in Figure~\ref{SIE_delens}, the posteriors converge toward the true value, while the credible intervals decrease as additional realizations are accumulated.
In contrast, for the SIS-delensing case shown in Figure~\ref{SIS_delens}, the bias is not apparent when only $N=10$ realizations are combined, but becomes increasingly evident as more realizations are accumulated.
These results show that lens-reconstruction biases, although potentially subtle in individual lensed dark-siren measurements, can accumulate and lead to a significant systematic offset in joint analyses, particularly within the \texttt{gwcosmo} framework.
\subsection{Selection bias in lensing identification}

In the lensed dark-siren case, the selection effects are more intricate than in the unlensed case, where detectability is determined solely by whether a signal exceeds the $\rho_{\mathrm{thr}}$ and only the population of dark sirens needs to be considered.
For lensed events, the selection effects must account for not only the $\rho_\mathrm{net}$ of individual lensed signals, but also for the requirement that the detected signals are correctly identified as originating from the same underlying GW source, quantified through the false alarm rate of the lensing hypothesis.

It is possible to misidentify two or more independent unlensed signals as lensed signals from the same source, or to misclassify a single lensed signal, whose counterpart has not yet arrived, is missed, or is too faint to be detected, as unlensed.
Such misidentification results in incorrect selection-bias corrections as well as a failure in lens reconstruction, leading to a biased recovery of the source luminosity distance. 
Consequently, the lensing misidentification can substantially bias the inferred $H_{0}$, particularly in analyses involving an increasing number of dark sirens as more events are observed.

Figure~\ref{fig:hidden_bias} illustrates the $H_0$ posteriors that are biased due to misidentification in the two scenarios described above. 
To isolate the effect of misidentification from uncertainties arising from incomplete catalogs, we use the complete UGC0 and LGC0 catalogs.
In the first scenario, we replace 4, 8, or 12 unlensed signals out of 250 with two, four, or six pairs of doubly-lensed signals that are intentionally misclassified as unlensed by construction to mimic the effect of pipeline misidentification.
We then estimate $H_0$ using UGC0 along with the selection correction appropriate for unlensed signals.
Figure~\ref{L_in_UL} shows the $H_0$ posteriors from the three cases, together with the reference $H_0$ posterior from the unlensed case with UGC0 in Figure~\ref{fig:final H0} for comparison.
The lensed signals remain undelensed, so their inferred luminosity distances are systematically biased, typically toward smaller values, as the magnification factor is generally greater than 1. 
This bias results in a contribution to higher $H_0$ values, as illustrated in Figure~\ref{no_delens}.
The impact of the misclassification is, however, subtle: for every increase of two in $N^{\rm DBL}{\rm pair}$, the median of the $H{0}$ posterior shifts upward by only about one unit, as the fraction of affected signals remains small relative to the total population.

In the second scenario, we replace one, two, or three doubly-lensed systems among the six doubles and two quads with the same number of independent unlensed signals, which are coincidentally identified as a lensed pair from the same GW source by the pipeline. 
The $H_0$ inference is then performed using LGC0 and the selection correction for lensed signals.
Figure~\ref{UL_in_L} shows the three cases, with the $H_0$ posterior from the lensed case with LGC0 in Figure~\ref{fig:final H0} included for reference.
Since the unlensed signals are erroneously delensed, their inferred distances are systematically larger than the true values, causing these signals to bias the $H_0$ posterior toward lower values.
Consequently, as the number of unlensed pairs increases, the $H_0$ posterior shifts progressively to lower values.
Compared to the first scenario, the fraction of misclassified signals in the analyzed population is larger, resulting in a more pronounced bias in $H_0$.
\section{Discussion}\label{sec:discussion}
We have shown that strong lensing substantially enhances the cosmological utility of dark sirens.
With a galaxy-galaxy lensing catalog, the improved sky localization and tighter luminosity-distance constraints provided by lensing enable significantly more informative redshift inference compared to the unlensed case.
We demonstrate that even a single strongly lensed dark siren, particularly a quadruply lensed event, can yield competitive constraints on $H_{0}$, approaching the precision of bright sirens. 
Furthermore, combining a few lensed dark sirens leads to a substantial improvement in the Hubble constant estimate, achieving constraints comparable to, or even tighter than, those obtained from a few hundred samples of unlensed events.

Robust cosmological inference from strongly lensed dark sirens requires accurate lens reconstruction to recover the intrinsic luminosity distance to the host galaxy.
With both EM and GW observations, lens reconstruction is generally reliable when the lens mass distribution is well described by a simple profile, such as the SIE profile adopted in this work.
Although the SIE model is a widely used and representative model for describing galactic mass profiles, real galaxies exhibit structural complexity beyond this idealization. 
Substructures and small-scale mass objects embedded within galaxies can introduce additional lensing effects.
Such effects may induce wave-optics signatures in GW signals, thereby modifying the observed waveform morphology~\citep{diego2019observational,PhysRevD.101.123512,cheung2021stellar,mishra2021gravitational,meena2022gravitational,seo2022improving,yeung2023detectability,shan2023microlensing}.
Accurate lens reconstruction in these regimes therefore requires incorporating wave-optics corrections associated with the small-scale structures.

Beyond the small-scale complexities, lens reconstruction for cosmological inference also suffers from a more fundamental large-scale degeneracy, the mass-sheet degeneracy (MSD)~\citep{falco1985model}.
The MSD arises because the observed lensed image positions and the relative lensing observables remain invariant under a transformation that rescales the projected mass density while introducing a uniform mass sheet.
As a consequence, lensing data alone cannot uniquely determine the absolute mass normalization or the time-delay distance, leading to a direct degeneracy with the inferred value of $H_{0}$.

The MSD can be broken either by multiple sources at different redshifts lensed by the same galaxy, which is extremely rare, or by introducing an independent physical constraint not determined solely by the lensing geometry.
A powerful tool is stellar kinematics, specifically through measurements of the velocity dispersion of stars bound to the lens galaxy~\citep{schneider2013mass}. 
These measurements probe the gravitational potential of the lens, thereby constraining its mass distribution independently of the lensing observables.
By jointly modeling the lensing configuration together with stellar kinematics, in particular the velocity dispersion of the lens galaxy, the mass normalization inferred from lensing, such as the projected mass enclosed within the Einstein radius, can be directly compared with that implied by dynamical measurements, effectively constraining the mass-sheet transformation parameter. 
Furthermore, in the wave-optics regime, distortions induced by substructure in the lens galaxy can provide additional constraints that help break the degeneracy~\citep{cremonese2021breaking}.

As discussed above, strongly lensed GW systems can provide tight constraints on $H_0$, even when their number is much smaller than that of unlensed events, and increasing the number of lensing detections to several tens of systems further tightens the precision to the few-percent level.
Next-generation GW detectors are expected to detect a large number of SLGWs ($\mathcal{O}(100)$ for a relative detection rate of lensed GWs equal to 1/1000), as the optical depth for target GW sources will increase significantly, and they may even detect demagnified lensed signals.
With improved sky localization and higher $\rho_{\mathrm{mean}}$, the identification and characterization of lensed GW sources will be significantly enhanced compared to current ground-based detectors~\citep{chen2026forecasting}.
Furthermore, space-based detectors like LISA~\citep{amaro2017laser} will achieve sub-degree-level localization even without lensing~\citep{babak2021lisa}. 
Strongly lensed supermassive black hole mergers and extreme mass-ratio inspirals will provide powerful probes for constraining $H_{0}$~\citep{toscani2024strongly}.
However, galaxy catalog incompleteness is expected to increase as the detector horizon extends to higher redshift. 
The impact of this incompleteness will therefore become increasingly significant, since the cosmological inference will be progressively dominated by prior assumptions in regions where the galaxy catalog is incomplete.
We leave the investigation of catalog incompleteness in the context of future detectors as future work.

In addition, since our work focuses exclusively on $H_{0}$, we do not consider hyperparameters describing the population properties of BBHs and lens galaxies, nor additional cosmological parameters such as the matter density parameter, $\Omega_{m}$, and the dark energy equation-of-state parameter, $w_{0}$.
Strong lensing, however, has the potential to provide complementary constraints on cosmological parameters beyond $H_{0}$~\citep{jana2023cosmography,chen2026forecasting}.
A fully hierarchical analysis that incorporates population-level parameters for both sources and lenses may yield a more reliable inference of $H_{0}$ as well as $\Omega_{m}$ and $w_{0}$.
In particular, realistic population models for lens galaxies, such as the halo mass function, redshift distribution, and structural and luminosity distributions, should be incorporated in future analysis.

In summary, we develop a comprehensive framework to infer $H_0$ using single and multiple lensed dark sirens, incorporating lensed galaxy catalogs and lensing selection effects, such as detection efficiency and catalog incompleteness.
We systematically quantify how the S/N of lensed GW signals, intrinsic source redshift, and delensing uncertainties affect the precision of recovered luminosity distances and the resulting $H_0$ posterior.
The analysis further demonstrates that even a small fraction of misclassified events can induce measurable systematic biases in $H_0$.
The results highlight the importance of accurate lens modeling, robust lens identification, and consistent treatment of selection effects in future GW cosmology involving strong lensing.
\section*{acknowledgments}
We thank Mick Wright for their valuable comments and feedback.
E.S. is supported by grants from the College of Science and Engineering of the University of Glasgow.
The work of K.K. is supported by the Korea Astronomy and Space Science Institute under the R\&D program (Project No.~2026-1-810-00) supervised by the Ministry of Science and ICT of the Korea Government.
Z.L. is supported by Chinese Scholarship Council.
J.J. acknowledges support from the FNRS.
R.G. and M.H. are supported by the Science and Technology Facilities Council (Grant Ref.~ST/V005634/1).
The authors are grateful for computational resources provided by the LIGO Laboratory and supported by the National Science Foundation Grants PHY-0757058 and PHY-0823459.
This material is based upon work supported by NSF's LIGO Laboratory which is a major facility fully funded by the National Science Foundation.
This work has made use of CosmoHub, developed by PIC (maintained by IFAE and CIEMAT) in collaboration with ICE-CSIC. It received funding from the Spanish government (grant EQC2021-007479-P funded by MCIN/AEI/10.13039/501100011033), the EU NextGeneration/PRTR (PRTR-C17.I1), and the Generalitat de Catalunya.

\newpage
\bibliography{main}{}  
\bibliographystyle{aasjournal}

\appendix
\section{Strong lensing optical depth}\label{app:optical_depth}
The probability that a source at redshift $z_{s}$ is strongly lensed for a given cosmology can be expressed as
\begin{equation}
\begin{aligned}
P_{\rm SL}(z_{s}) &= 1-e^{-\tau(z_{s})} \\
&\simeq \tau(z_{s}) 
\end{aligned}
\end{equation}
where $\tau(z_{s})$ denotes the optical depth, generally satisfying $\tau \ll 1$ in realistic galaxy-galaxy lensing.
Quantitatively, the optical depth corresponds to the fraction of the sky over which a background source is strongly lensed by foreground objects. It is formally defined as the line-of-sight integral of the lens number density weighted by their individual lensing cross-sections.
With a given $H_{0}$ and cosmological parameters, $\tau(z_{s})$ can be written as
\begin{equation}
\label{tau_general}
    \tau(z_{s}) = \int^{z_{s}}_{0} dz_{l} \frac{dV_{c}}{\delta \Omega dz_{l}} \int  d\theta_{l} n_{l}(z_{l},\theta_{l})\hat{\sigma}_{\rm SL}(z_{s},z_{l},\theta_{l}) 
\end{equation}
where $dV_{c}/ (\delta \Omega dz_{l})$ denotes the differential comoving volume ($V_{c}$) per steradian ($\Omega$) at lens redshift $z_{l}$, $n_{l}$ is the comoving number density of lenses with properties $\theta_{l}$, and $\hat{\sigma}_{\rm SL}$ represents the strong lensing cross-section of an individual lens for a given $z_s$, $z_{l}$ and $\theta_{l}$.
Here, since we adopt the SIE profile as our lens model, the lens parameters relevant for computing the strong lensing cross-section are the axis-ratio and velocity dispersion, that is, $\theta_{l}=(q,\sigma_{v})$.

Since the SIE model produces either double- or quadruple-image systems, we separately consider the cross-section term $\hat{\sigma}_\mathrm{SL}$ of Eq.~\eqref{tau_general} as $\hat{\sigma}_{\rm dbl}$ and $\hat{\sigma}_{\rm quad}$ for each system, respectively.
For galaxies with axis ratios above $q \gtrsim 0.394$, the corresponding cross-sections, $\hat{\sigma}_{\rm dbl}$ and $\hat{\sigma}_{\rm dbl}$, can be expressed analytically as
\begin{equation}
    \hat{\sigma}_{\rm dbl}  = r^{2}_{\rm E}(z_{s},z_{l},\sigma_{v}) \hat{\mathcal{A}}_{\rm cut}(q); \quad 
    \hat{\sigma}_{\rm quad} = r^{2}_{\rm E}(z_{s},z_{l},\sigma_{v}) \hat{\mathcal{A}}_{\rm caustic}(q),
\end{equation}
where $r_{\rm E}$~\citep{kormann1994isothermal} denotes the Einstein radius, and $\hat{\mathcal{A}}_{\rm cut}$ and $\hat{\mathcal{A}}_{\rm caustic}$ denote the areas enclosed by the radial and tangential caustics of an SIE lens\footnote{For an isothermal profile, a radial caustic turns into a \emph{cut}, on which a source does not produce infinitely magnified images.}, respectively.
For more elliptical systems ($q < 0.394$), where the analytic forms become inaccurate, we numerically calculated the cross-sections using the ray-shooting technique.

The comoving number density $n_{l}(z_{l},\theta_{l})$, appearing in Eq.~\eqref{tau_general}, represents the number of lenses per differential comoving volume and per differential interval in $z_{l}$ and $\theta_{l}$ as described above. 
Based on this, we now define $N_{\rm tot}$ as the total number of galaxies in the Universe, distributed in lens redshift, axis ratio, and velocity dispersion.
Substituting the associated number into Eq.~\eqref{tau_general}, the optical depth can be written as
\begin{equation}
    \tau(z_{s}) = \int^{z_{s}}_{0} dz_{l} \frac{dV_{c}}{\delta \Omega dz_{l}}\iint   dq d\sigma_{v} \frac{N_{\rm tot}}{dV_{c}dz_{l}dqd\sigma_{v}}r^{2}_{\rm E}(z_{s},z_{l},\sigma_{v}) \hat{\mathcal{A}}_{\rm cut \; or \; caustic}(q)  ,
\end{equation}
which provides a convenient form for computing the optical depth.
\section{Source and lens populations}\label{app:populations}
In this section, we describe the population of sources, i.e., the component masses and redshifts of BBHs, as well as the population of lenses, i.e., 
galaxies that are described as SIE, where the lens parameters include the velocity dispersion and axis ratio.
\subsection{Component mass}
The primary mass, $m_1$, in the mock BBH population is drawn from a \textsc{Powerlaw+Peak} model, comprising a truncated power-law component defined over $m_{\rm min}= 5M_{\odot}$ and $m_{\rm max}=112M_{\odot}$, with additional tapering applied at the low-mass end and an added Gaussian component~\citep{abbott2023population}.
Within the mass range, the intrinsic distribution of $m_1$ is given by
\begin{equation}
\label{eq:m1_pop}
p_{\rm pop}\left(m_1 \mid  \alpha, \lambda_{\mathrm{g}}, \mu_{\mathrm{g}}, \sigma_{\mathrm{g}}, m_{\rm min}, \delta_{m}\right)=\left[\left(1-\lambda_{\mathrm{g}}\right) m^{-\alpha}_{1}+ 
\frac{\lambda_{\mathrm{g}}}{\sqrt{2\pi}\,\sigma_g}
\exp\left(
-\frac{(m_1 - \mu_g)^2}{2\sigma_g^2}
\right)\right]S(m_{1}|m_{\rm min}, \delta_{m}),
\end{equation}
where $\alpha$ denotes the power-law index, and $\lambda_{g}$, $\mu_{g}$, and $\sigma_{g}$ correspond to the mixture fraction, mean, and standard deviation of the Gaussian component, respectively.
$S(m_{1}|m_{\rm min},\delta_{m})$ represents a smoothing function, which is given by
\begin{equation}
    \label{eq:smoothing}
    \begin{aligned}
& S\left(m \mid m_{\min }, \delta_m\right) \\
& \quad= \begin{cases}0 & \left(m<m_{\min }\right), \\
{\left[\mathrm{exp}\left(\frac{\delta_{m}}{m-m_{\rm min}}+\frac{\delta_{m}}{m-m_{\rm{min}}-\delta_{m}}\right)+1\right]^{-1}} & \left(m_{\min } \leq m<m_{\min }+\delta_m\right), \\
1 & \left(m \geq m_{\min }+\delta_m\right),\end{cases}
\end{aligned}
\end{equation}
where $\delta_{m}$ sets the width of the low-mass taper applied near $m_{\rm min}$.

For secondary mass $m_{2}$, rather than sampling it directly, we draw the mass ratio $q_{m} \equiv m_{2}/m_{1}$ from a truncated power-law distribution, with the low-end smoothing function defined in Eq.~\eqref{eq:smoothing}.
The intrinsic distribution of the mass-ratio is given by
\begin{equation}
\label{eq:qm_pop}
    p_{\rm pop}\left(q_{m} \mid m_{\min }, m_{1}, \beta \right)  = (q_{m})^{\beta} S(m_{2}|m_{\rm min},\delta_{m}) \quad \text{when} \;\; m_{\rm min} < m_{2} < m_{1}.
\end{equation}
where $\beta$ denotes the power-law index of the mass-ratio distribution.
We adopt the hyperparameter values inferred from the GWTC-3 analysis~\citep{abbott2023population}, as summarized in Table~\ref{tab:hyperparameters}.

\begin{table}[h]
\centering
\begin{tabular}{l l l}
\hline\hline
Parameter & Description & Injected value \\
\hline
$\alpha$      & Spectral index for the power law of the primary mass distribution & 3.78 \\
$\beta$     & Spectral index for the power law of the mass ratio distribution & 0.81 \\
$m_{\min}$    & Minimum mass of the power-law component of the black hole mass distribution & 4.98$M_{\odot}$ \\
$m_{\max}$    & Maximum mass of the power-law component of the black hole mass distribution & 112.5$M_{\odot}$ \\
$\lambda_{\rm g}$ & Fraction of BBH systems in the Gaussian component & 0.03 \\
$\mu_{g}$  & Mean of the Gaussian component in the primary mass distribution & 32.27$M_\odot$ \\
$\sigma_{g}$    & Standard deviation of the Gaussian component in the primary mass distribution & 3.88$M_\odot$ \\
$\delta_m$    & Range of mass tapering at the lower end of the mass distribution & 4.8$M_{\odot}$ \\
$z_{p}$    & Redshift at which the distribution peaks & 2.7 \\
$\gamma$   & Power-law index of the redshift distribution for $z < z_p$ & 4.59 \\
$\kappa$    &  Power-law index of the redshift distribution for $z > z_p$& 2.86 \\
\hline
\end{tabular}
\caption{List of hyperparameters of the BBH population model and their injected values used in our simiulations.}
\label{tab:hyperparameters}
\end{table}
\subsection{Merger rate and redshift evolution}
The redshift distribution of the BBHs depends on their merger rate in the given cosmology.
The number of compact binary mergers per unit time and unit redshift is given by
\begin{equation}
\label{eq:diff_merger_rate}
    \frac{d^{2}N}{dtdz} = \frac{\mathcal{R}(z)}{1+z}\frac{\partial V_{c}}{\partial z},
\end{equation}
where $\mathcal{R}(z)$ is the merger rate at redshift $z$.
Using the functional form of the star formation rate from \citet{madau2014cosmic} and the parametrization in \citet{callister2020shouts}, the merger rate in the detector frame can be expressed as
\begin{equation}
\label{eq:z_pop}
\mathcal{R}(z) = \mathcal{R}_0(1+z)^\gamma \frac{1+\left(1+z_p\right)^{-(\gamma+\kappa)}}{1+\left(\frac{1+z}{1+z_p}\right)^{\gamma+\kappa}},
\end{equation}
where $R_{0}$ is the local merger rate, $z_{p}$ is
the redshift at which the merger-rate density peaks, and $\gamma$ and $\kappa$ denote the power-law indices of the redshift distribution before and after $z_{p}$, respectively.
In our simulation, we adopt $R_{0}=20~\mathrm{Gpc^{-3}yr^{-1}}$, and the values for the other parameters are given in Table~\ref{tab:hyperparameters}. 
\subsection{Velocity dispersion and axis-ratio of galaxies}\label{app:lens_params}
As mentioned above, the \texttt{MICECATv2} catalog does not provide galaxy velocity dispersions or axis ratios.
We therefore sample these parameters from the distributions adopted in \citet{xu2022please}.
For the velocity dispersion $\sigma_{v}$, we draw it from a Schechter function, which is given by
\begin{equation}
\phi\left(\sigma_{v} \mid z_l\right)=\phi_*\left(z_l\right)\left(\frac{\sigma_{v}}{\sigma_*}\right)^{\alpha_{\mathrm{g}}} e^{-\left(\frac{\sigma_{v}}{\sigma_*}\right)^{\beta_{\mathrm{g}}}} \frac{\beta_{\mathrm{g}}}{\Gamma\left(\alpha_{\mathrm{g}} / \beta_{\mathrm{g}}\right)} \frac{1}{\sigma_{v}},
\end{equation}
where $\phi_{*}(z_{l})$ is the number density of galaxies at the redshift $z_{l}$ in the catalog and provides the normalization of the distribution.
$\sigma_{*}$ is the characteristic velocity dispersion representing typical massive elliptical galaxies,
$\alpha_{g}$ is the power-law index governing the low-$\sigma_{v}$, and $\beta_{g}$ controls the steepness of the exponential decline at high-$\sigma_{v}$ end.
In this work, we adopt $\sigma_{*}=161\mathrm{km\cdot s^{-1}}$  $\alpha_{\rm g}=2.32$, and $\beta_{\rm g}=2.67$, as measured by \citet{choi2007internal}.

The galaxy axis ratios, $q$, are sampled from a Rayleigh distribution, whose scale parameter $s_{g}$ depends on the velocity dispersion of the galaxy, as given by
\begin{equation}
\label{rayleigh}
p_{\rm pop}\left(q \mid s_{g}=A+B \sigma_{v}\right)=\frac{1-q}{s_{g}^2} \exp \left[\frac{-\left(1-q\right)^2}{2 s_{g}^2}\right],
\end{equation}
where $A=0.38$ and $B=5.7\times 10^{-4}[\mathrm{km\cdot s^{-1}}]^{-1}$~\citep{collett2015population,xu2022please}.
Eq.~\eqref{rayleigh} indicates that galaxies with larger $\sigma_{v}$ tend to have higher $q$ values.
\section{Hubble constant posteriors}\label{app:H0posterior}
Following the formalism of \citet{gray2023joint}, for a set of lensed dark sirens, ${x_{\rm lgw}}$, with corresponding detection parameters, ${D_{\rm lgw}}$ (where each element is either $0$ or $1$), and assuming $N_{\rm lens,det}$ detected lensed systems, the posterior of the Hubble constant, $H_{0}$, can be written as
\begin{equation}\label{lgw_h0_pos}
\begin{aligned}
&p(H_{0} | \{x_{\rm lgw}\}, \{D_{\rm lgw}\}) \propto p(H_{0}) p(N_{\rm lens, det} | H_{0})\\
&\times \left[ \iiiint  p(D_{\rm lgw} |  \theta_{s}, \theta_l, z_{s}, z_{l}, H_{0}) p(\theta_s | H_{0}) \sum_{j}^{N_{\rm pix}} p(z_{s},z_{l}, \theta_l | \Omega_j, H_{0},s,l) d\theta_l d\theta_{s} dz_{l} dz_{s} \right]^{-N_{\rm lens,det}}\\
&\times \prod_{i}^{N_{\rm lens, det}} \left[\iiiint  \sum_{j}^{N_{\rm pix}} p(x_{{\rm lgw}, i} | \Omega_j, \theta_{s},\theta_l, z_{s}, z_{l} ,H_{0}) p(\theta_{s} | H_{0}) p(z_{s},z_{l}, \theta_l | \Omega_j, H_{0},s,l) d\theta_l d\theta_{s} dz_{l}dz_{s} \right]
\end{aligned}
\end{equation}
where $\theta_{s}$, $\theta_{l}$, $z_{s}$, and $z_{l}$ denote the GW source parameters excluding the distance, the lens parameters of the source-lens galaxy system, the source redshift, and the lens galaxy redshift, respectively.
$p(z_{s},z_{l} | \Omega_j, H_{0},s)$ is the LoS redshift distribution of source and lens galaxies obtained from the galaxy-galaxy lensing catalog.
The GW sky localization is discretized into $N_{\rm pix}$ pixels, with $\Omega_{j}$ corresponding to the sky direction of the $j^{\rm th}$ pixel.
The indicator $s$ specifies whether a given source galaxy in a lensing system hosts a GW source.
$p(H_{0})$ denotes the prior on the Hubble constant, which is a uniform distribution over the range of $[20,140]~\rm{km \cdot s^{-1}\cdot Mpc^{-1}}$.
More detailed descriptions of the remaining terms are provided in the following subsections.
\subsection{Selection effect associated with lensed gravitational waves}\label{app:selection}
Selection bias arises when the number of detectable signals differs from the total number of mergers due to limited detector sensitivity. 
The inferred $H_{0}$ becomes biased if selection effects are ignored. 
For example, low-redshift events are more easily detected, which would incorrectly suggest an overabundance of GW sources at low redshift. 
To correct for this, the likelihood of each GW event must be divided by its detection efficiency, that is, the probability that an event with given parameters would be observed.

Given the detection probability for a source with parameters $\vec{\theta}= (\theta_{s}, \theta_l, z_{s}, z_{l})$, $p_{\rm det}(\vec{\theta})$ and the underlying population distribution $p_{\rm pop}(\vec{\theta})$ for a given set of hyperparameters $\Lambda$, such as Eqs.~\eqref{eq:m1_pop}, \eqref{eq:qm_pop}, and \eqref{eq:z_pop}, the expected number of detections of lensed GW events is given by
\begin{equation}
\begin{aligned}
\label{eq:expected_lens_number}
\langle N_{\rm lens,det}\left(\Lambda\right) \rangle &=  N_{\rm lens}\int p_{\rm det} (\vec{\theta}) p_{\rm pop}(\vec{\theta}| \Lambda) d \vec{\theta} \\
 &= N_{\rm lens} \iiiint p(D_{\rm lgw} | \theta_{s}, \theta_l, z_{s}, z_{l},  \Lambda) p(\theta_{s}, \theta_l, z_{s}, z_{l} | \Lambda) d \theta_{s} d \theta_{l}  d z_{l} d z_{s}
\end{aligned}
\end{equation}
where $N_{\rm lens}$ denotes the total number of lensed mergers~\citep{mandel2019extracting}.
The detection efficiency is the same as the ratio between the total and expected number of detection.
To efficiently evaluate the detection efficiency of lensed GW events, we generated numerous injections from a fiducial distribution $\pi(\vec{\theta})$ (see \citet{farr2019accuracy} for more details).
For a given set of detected lensed GW events, the detection efficiency, equivalent to the selection-effect term in Eq.~\eqref{lgw_h0_pos}, can be evaluated as follows
\begin{equation}\label{inj_for_selection}
\begin{aligned}
    \frac{\langle N_{\rm lens,det}\left(\Lambda\right) \rangle}{N_{\rm lens}} &= \iiiint p(D_{\rm lgw} | \theta_{s}, \theta_l, z_{s}, z_{l},  \Lambda) p(\theta_{s}, \theta_l, z_{s}, z_{l} | \Lambda) d \theta_{s} d \theta_{l}  d z_{l} d z_{s} \\
    & \simeq \frac{1}{N_{\rm inj}}\sum^{N_{\rm inj,det}}_{i} \frac{p_{\rm pop}(\vec{\theta_{i}}|\Lambda)}{\pi(\vec{\theta}_{i})} \\
    & = \frac{1}{N_{\rm inj}}\sum^{N_{\rm inj,det}}_{i} \frac{p_{\rm pop}(m_{1,i},m_{2,i},z_{s,i}|\Lambda)}{\pi(m_{1,i},m_{2,i},z_{s,i})} \frac{p_{\rm pop}(\theta_{l,i},z_{l,i}|z_{s,i}\Lambda)}{\pi(\theta_{l,i},z_{l,i})} \\
    & = \frac{1}{N_{\rm inj}}\sum^{N_{\rm inj,det}}_{i} \frac{p_{\rm pop}(m_{1,i},m_{2,i},z_{s,i}|\Lambda)}{\pi(m_{1,i},m_{2,i},z_{s,i})} \frac{p_{\rm pop}(\vec{\mu}_{i}|\Lambda)}{\pi(\vec{\mu}_{i})} 
\end{aligned}
\end{equation}
where $N_{\rm inj}$ is the total number of injections, $N_{\rm inj,det}$ is the number of detected injections, and $\vec{\mu}_{i}$ are the individual magnification factors of $i^{\rm th}$ lensing system.
Instead of sampling the $\theta_{l}$ and $z_l$ directly, we first compute $\vec{\mu}$ from the lens parameters, creating multiple images under a specific lens model. 
The resulting magnification factors define a distribution from which we draw samples, as these factors directly affect the S/N of the lensed signals.
We adopt the population models in Appendix~\ref{app:populations} for the source-frame primary and secondary masses and redshift. 
All other source parameters are drawn from hyperparameter-independent distributions.
For these latter parameters, the population prior $p_{\rm pop}$ reduces to the prior $\pi$ and is thus omitted.

Since detectability is determined in the detector frame, we draw injections in the detector frame, which modifies Eq.~\eqref{inj_for_selection} as follows,
\begin{equation}\label{eq:inj_det_frame}
\begin{aligned}
    \frac{\langle N_{\rm lens,det}\left(\Lambda\right) \rangle}{N_{\rm lens}} & \simeq \frac{1}{N_{\rm inj}}\sum^{N_{\rm inj,det}}_{i} \frac{p_{\rm pop}(m_{1,i},m_{2,i},z_{s,i}|\Lambda)}{\pi(m_{1,i},m_{2,i},z_{s,i})} \frac{p_{\rm pop}(\vec{\mu}_{i}|\Lambda)}{\pi(\vec{\mu}_{i})} \\
    &= \frac{1}{N_{\rm inj}}\sum^{N_{\rm inj,det}}_{i} \frac{p_{\rm pop}(m_{1,i},m_{2,i},z_{s,i},\vec{\mu}_{i}|\Lambda)}{\pi(m^{d}_{1,i},m^{d}_{2,i},\vec{d^{\rm eff}_{L,i}})} \left| \frac{\partial(m_{1,i},m_{2,i},z_{s,i},\vec{\mu}_{i})}{\partial(m^{d}_{1,i},m^{d}_{2,i},\vec{d^{\rm eff}_{L,i}})}\right|\\
    & =  \frac{1}{N_{\rm inj}}\sum^{N_{\rm inj,det}}_{i} \frac{p_{\rm pop}(m_{1,i},m_{2,i},z_{s,i},\vec{\mu}_{i}|\Lambda)}{\pi(m^{d}_{1,i},m^{d}_{2,i},\vec{d^{\rm eff}_{L,i}})}
    \prod_{j} \left| \frac{2\mu_{i,j}}{d^{\rm eff}_{L,i,j}}\right| \frac{1}{(1+z_{s,i})^{2}}\left|\frac{\partial z_{s}}{\partial d_{L}}\right|_{z_{s}=z_{s,i}},
\end{aligned}
\end{equation}
where $m^{d}_{1}$ and $m^{d}_{2}$ are detector-frame primary and secondary masses.
Although injections are generated in the detector frame, the selection function is evaluated by reweighting them to the source-frame population model, including the full Jacobian associated with the transformation between $(m^{d}_{1},m^{d}_{2},d^{\rm eff}_{L})$ and $(m_{1},m_{2},z_{s},\vec{\mu})$.

To determine whether a signal is detected, the unlensed case is straightforward. 
A single event is considered detected if its network S/N exceeds the chosen threshold, $\rho_{\rm net} = 8$.
However, for lensed events, multiple signals from the same source must not only exceed the network S/N threshold, but also be confirmed to originate from the same source and truly lensed through a lensing hypothesis test.
Strong lensing can be verified using JPE pipelines, such as \citet{liu2021identifying,janquart2021fast, janquart2023return,lo2023bayesian}.
In this study, we assume that all lensed signals pass the hypothesis test. 
Hence, the detection criterion reduces to verifying that lensed signals are detected above the S/N threshold and correctly identified as lensed.
Accordingly, $N_{\rm lens,det}$ in Eq.~\eqref{inj_for_selection} denotes the number of GW lensing systems that satisfy both the detection threshold and the lensing hypothesis test.

For doubly lensed GW systems, if one of the two signals does not exceed the detection threshold, it is not possible to determine from the detected signal alone whether the system is lensed.
For quadruply lensed systems, only two or three of the four images may be detected in realistic scenarios.
We impose the analogous requirement to the doubly-lensed case that at least two signals must exceed the detection threshold for the system to be identified as strongly lensed.
Since the lensing hypothesis test determines the number of detected lensed signals in each system, we consider three cases: two, three, or all four images detected, when accounting for selection effects.
It is important to note that partially detecting only two or three of the four images can make it initially unclear whether the system originates from a quadruply lensed source.
Accounting for selection effects under the assumption that such systems are doubly or triply lensed leads to a biased inference of $H_{0}$.

Nevertheless, adopting a specific lens model fixes the possible number of lensed signals and determines the population distributions of lensing observables, such as relative magnification factors and time delays.
Under the SIE model, for example, only two or four images can be produced.
Thus, if three images are detected, the event can be unambiguously identified as a quadruply-lensed system, allowing the appropriate quadruple-image selection effects to be applied.
When only two images are detected, however, both the doubly-lensed and quadruply-lensed interpretations remain possible, and selection effects from both cases must be considered.

The term $p_{\rm pop}(\vec{\theta}|\Lambda)$ in Eq.~\eqref{eq:expected_lens_number} can be expressed in terms of the population distribution function ($dN/d\vec{\theta}$)~\citep{mandel2019extracting,farr2019accuracy}.
From Eq.~\eqref{eq:diff_merger_rate}, $dN/d\vec{\theta}$ can then be related to the intrinsic merger rate, together with the corresponding optical depth as follows,
\begin{equation}
\label{eq:expected_Nlens}
\begin{aligned}
     \langle N_{\rm lens,det}\left(H_0\right) \rangle &= N_{\rm lens}\iiiint p(D_{\rm lgw} | \theta_{s}, \theta_l, z_{s}, z_{l},  H_{0}) \frac{1}{N_{\rm lens}} \left(\frac{d^{4}N_{\rm lens}}{d \theta_{s} d \theta_{l}  d z_{l} d z_{s}}\right)  d \theta_{s} d \theta_{l}  d z_{l} d z_{s} \\
     &=T_{\text {obs }}\iiiint p(D_{\rm lgw} | \theta_{s}, \theta_l, z_{s}, z_{l},  H_{0}) \tau(z_{s},H_{0})\frac{R(z_{s})}{1+z_{s}}\frac{\partial V_{c}}{\partial z_{s}}  d \theta_{s} d \theta_{l}  d z_{l} d z_{s},
\end{aligned}
\end{equation}
where $T_{\rm obs}$ denotes the observation time and $\tau(z_{s}, H_{0})$ is the optical depth, identical to that in Eq.~\eqref{tau_general}.
It encodes the distribution of $\theta_{l}$ and $z_{l}$, both of which depend on $H_{0}$ in contrast to the $z_{s}$ and $\theta_{s}$, which are independent of $H_{0}$.
Therefore, unlike in the unlensed case, the probability of detecting $N_{\rm lens,det}$ lensed dark siren system for a given value of $H_{0}$, $p(N_{\rm lens,det} \mid H_{0})$, cannot be neglected even when adopting a scale-invariant prior on the merger rate, $P(R) \propto 1/R$.
For a given expected number of detections ($\langle N_{\rm lens,det}\rangle$), $p(N_{\rm lens,det} \mid H_{0})$ in Eq.~\eqref{lgw_h0_pos} can be expressed as a Poisson distribution as follows,
\begin{equation}
p(N_{\rm lens,det} \mid H_{0})= e^{-\langle N_{\rm lens,det}\rangle} \left(\langle N_{\rm lens,det}\rangle \right)^{N_{\rm lens,det}}.
\end{equation}
\subsection{Line-of-sight redshift prior}\label{app:los}
The LoS redshift prior is constructed by using the measured redshift, luminosity, and apparent magnitude information of galaxies in a given catalog to identify which galaxies lie within the GW sky localization, to weight them by their probability of hosting a BBH merger (e.g., luminosity weighting) and being lensed, and to correct for the incompleteness of the galaxy catalog.
This framework enables the inference of the redshift distribution of the observed BBH events from GW observations.

For the purpose of identifying galaxies within a given GW localization, a galaxy-galaxy lensing catalog, in which both source and lens galaxy information is available for each system, can be more informative than a standard galaxy catalog containing only unlensed galaxies.
However, luminosity weighting and corrections for catalog incompleteness require the intrinsic source-galaxy luminosity (i.e., the absolute magnitude) and the redshift.
While galaxy-scale gravitational lensing is achromatic and leaves the observed redshift unchanged to within the precision of spectroscopic measurements, the absolute magnitude can be significantly altered by lensing magnification.
Therefore, delensing is required to recover the intrinsic absolute magnitudes of source galaxies in order to obtain robust luminosity weights and incompleteness corrections. 
Accordingly, the LoS redshift prior must account for whether both the source and lens galaxies are included in the catalog.

\subsubsection{Complete catalog}~\label{app:los_complete}
We first consider the complete catalog case, in which there is no apparent magnitude limit, such that every lens galaxy and all lensed images of the source galaxy for each system are included in the catalog.
In this case, the LoS redshift prior in eq.~\eqref{lgw_h0_pos}, $p(z_{s},z_{l},\theta_{l} | \Omega_j, H_{0}, s)$, is given by
\begin{equation}
\begin{aligned}
\label{LoS_complete}
p\left(z_s, z_l, \theta_l \mid \Omega_j, H_0, s, l \right) 
&= \iint  
p\left(z_s, z_l, \theta_l, M_{s}, \vec{m}'_{s},M_{l}, m_{l} \mid \Omega_j, H_0, s, l \right) \, dM_{s} \, d\vec{m}'_{s}, dM_{l} , dm_{l},
\end{aligned}
\end{equation}
where $M_{s}$ and $ M_{l}$ denote the intrinsic absolute magnitudes of source and lens galaxies, respectively.
$m_{l}$ and $\vec{m}'_{s}$ are the apparent magnitude of the lens galaxy and lensed source-galaxy images (see Eq.~\eqref{eq:lensed_observables}), respectively.
Note that $M_{s}$ cannot be correctly obtained from $\vec{m}'_{s}$ using the standard distance modulus alone, as the magnification factor of the lensed image must be taken into account.
Hereafter, we define $\vec{M} = (M_s, M_l)$ and $\vec{m} = (\vec{m}'_s, m_l)$ to concisely represent the absolute and apparent magnitudes.
The integrand in Eq.~\eqref{LoS_complete} can then be written as
\begin{equation}
\begin{aligned}
\label{eq:complete_catalog}
&p\left(z_{s},z_{l},\theta_{l} , \vec{M}, \vec{m} \mid \Omega_j, H_{0}, s, l, \mathcal{M}_{\rm SIE}\right)  \\
&= \frac{1}{p\left(s, l \mid  \Omega_j, H_{0} \right)} p\left(z_{s},z_{l},\theta_{l} , \vec{M}, \vec{m} \mid \Omega_j, H_{0}, \mathcal{M}_{\rm SIE}\right) p\left(s,l \mid z_{s},z_{l},\theta_{l} , \vec{M}, \Omega_j, H_{0}, \mathcal{M}_{\rm SIE} \right) \\
&=  \frac{1}{p\left(s,l \mid  \Omega_j, H_{0} \right)}\delta(M_{s}-M_{s}(z_{s},z_{l},\theta_{l},\vec{m}'_{s},H_{0}, \mathcal{M}_{\rm SIE})) \delta(M_{l}-M_{l}(z_{l},m_{l},H_{0})) \\
&\times p\left(z_{s},z_{l},\theta_{l}, \vec{m}'_{s}, m_{l}\mid  \Omega_j, H_{0}\right) p\left(s,l \mid z_{s},z_{l},\theta_{l}, M_{s},H_{0} \right),
\end{aligned} 
\end{equation}
where $\mathcal{M}_{\rm SIE}$ indicates that SIE profiles are assumed to infer $M_{s}$ from $\vec{m}'_{s}$ of their images.

Since all lensed images and the lens galaxy are observed in a complete catalog, the intrinsic luminosity of the source galaxy can, in principle, be accurately recovered. 
We therefore assume that the intrinsic absolute magnitude of the source galaxy is uniquely determined once the other measured parameters are specified, such that its prior can be represented as a delta function.
The absolute magnitude of the lens galaxy, $M_{l}$, is unaffected by lensing and independent of the source properties, and thus its prior can likewise be treated as a delta function\footnote{Note that, however, in galaxy-galaxy lensing systems, the spatial overlap between foreground lenses and background lensed source-galaxy images leads to blended photometry, which can bias photometric redshift estimates. \jj{Maybe add a ref to this}}.
Moreover, the probability that a galaxy hosts a GW source and that the detected lensed images are genuinely associated does not depend on $M_{l}$, $\Omega_{j}$, or $\mathcal{M}_{\rm SIE}$, which are omitted in the last term $p(s,l|...)$.

The LoS redshift prior for the complete catalog case can be written as
\begin{equation}
\begin{aligned}
\label{incat}
&p\left(z_s, z_l, \theta_l \mid \Omega_j, H_0, s,l, \mathcal{M}_{\rm SIE} \right)  \\
&= \frac{1}{p\left(s,l \mid  \Omega_j, H_{0} \right)} \frac{1}{N_{\rm Lgal}(\Omega_{j})}\sum^{N_{\rm Lgal}}_{k} p\left(z_{s},z_{l}, \theta_{l} \mid \hat{z}_{s,k}, \hat{z}_{l,k} ,\hat{\theta}_{l,k} \right) 
p\left(s,l \mid z_{s}, z_{l},\theta_{l}, M_{s}(z_{s},z_{l},\theta_{l},\hat{m}'_{s,k},H_{0}, \mathcal{M}_{\rm SIE}) \right),
\end{aligned}
\end{equation}
where $N_{\rm Lgal}$ denotes the number of lensing systems within the sky patch $\Omega_j$. 
Following the notation of \cite{gray2023joint}, the hat notation denotes the measured value for the $k^{\rm th}$ galaxy-galaxy lensing system, and the corresponding $p(z_s,z_l, \theta_l)$ are modeled as Gaussian distributions centered on these measured values.
For $z_s$ and $z_l$, we adopt the spectroscopic uncertainties used in Sec.~\ref{subsec:HLlaw}, and for the lens parameters we assume a standard deviation of 0.01, which corresponds to the grid resolution used in the delensing procedure described in~\cite{seo2024inferring}.
Note that for certain lensing parameters, such as the impact parameter, which must be inferred from the lensing geometry rather than measured directly, as is the case for the axis ratio from photometry, the Gaussian assumption represents a simplification.
As shown in Eq.~\eqref{incat}, the magnitudes of the lens galaxies are not included, which can be neglected when deriving the LoS redshift priors, and thus we will omit $M_{l}$ and $m_{l}$ and not explicitly write them in the following.

Since no explicit correlation between GW source redshift and host galaxy luminosity is known, the term $p(s,l|...)$ can be factorized as $p(s,l|z_{s},z_{l},\theta_{l},M_{s},H_{0}) \propto p(s|z_{s})p(s|M_{s})p(l|s,z_{s},z_{l},\theta_{l},H_{0})$.
For simplicity, one may reduce the complexity of the LoS computation by disabling the luminosity weighting, i.e., setting $p(s|M_{s}) \propto const.$.
In this case, all source galaxies are assumed to have equal probability of hosting a BBH merger, and the resulting LoS redshift prior depends only on the merger rate as a function of redshift and the distribution of galaxies.
\subsubsection{Incomplete catalog}~\label{app:los_incomplete}
In practice, galaxy catalogs are often incomplete due to observational limitations.
In this section, we examine the impact of an apparent magnitude threshold $m_{\rm th}$, which may cause certain lens galaxies or lensed source-galaxy images to be missing from the catalog.
Similar to the unlensed case, in which inclusion in the catalog is determined solely by whether a galaxy is sufficiently bright to be detected, the lensed case additionally requires accounting for whether a system is identified as strongly lensed ($GG$) or not ($\overline{GG}$).
With these definitions, Eq.~\eqref{LoS_complete} can be rewritten as
\begin{equation}
\begin{aligned}
\label{LoS_redshift_lensgalaxy}
p\left(z_s, z_l, \theta_l \mid \Omega_j, H_0, s, \mathcal{M}_{\rm SIE} \right) 
&= \iint \sum_{g = GG, \overline{GG}}
p\left(z_s, z_l, \theta_l, M_{s}, \vec{m}'_{s}, g \mid \Omega_j, H_0, s,l, \mathcal{M}_{\rm SIE}\right) \, dM_{s} d\vec{m}'_{s} \\
&= \sum_{g = GG, \overline{GG}} 
p\left(g \mid \Omega_j, H_0, s,l, \mathcal{M}_{\rm SIE} \right) 
\iint p\left(z_s, z_l, \theta_l, M_{s}, \vec{m}'_{s} \mid g, \Omega_j, H_0, s,l,\mathcal{M}_{\rm SIE} \right) \, dM_{s} d\vec{m}'_{s}.
\end{aligned}
\end{equation}
The identification of a galaxy-galaxy strong-lensing system depends on the detection of multiple lensed source-galaxy images, independent of the visibility of the lens galaxy itself.
However, recovering the intrinsic absolute magnitude $M_s$, required for accurate luminosity weighting, is necessary for a delensing procedure.
To assess the impact of imperfect delensing, we consider six possibilities based on the visibility of the lens galaxy and the source-galaxy images: whether the lens is in- or out-of-catalog, and whether all, a subset, or none of the source images are in the catalog (see Table~\ref{tab:los_terms} for details).

\begin{table*}[hb]
\centering
\begin{tabular}{c c c c c}
\hline\hline
Case & Catalog & $z_s, \vec{m}'_{s}$ & $z_l, \theta_{l}$ & Delensing \\
\hline
$(G,LG)$ & In & \cmark & \cmark &  Very accurate (0-5\%)\\
$(G',LG)$ & In & \cmark & \cmark &  Accurate (5-20\%) \\
$(G,\overline{LG})$ & In & \cmark & \xmark & Possible (20-50\%)\\
$(G',\overline{LG})$ & Mixed & \cmark & \xmark &  Poor (50\%-) or Impossible \\
$(\bar{G},LG)$ & Out & \xmark & \xmark &  Impossible \\
$(\bar{G},\overline{LG})$ & Out & \xmark &  \xmark & Impossible  \\
\hline
\end{tabular}
\caption{Information availability for the six possibilities determined by the observational accessibility of the lens galaxy and the lensed source-galaxy images.
$G$ denotes the case in which all lensed images are included in the catalog, $\bar{G}$ denotes the case in which none of the images are included, and $G'$ denotes the case in which only a subset of the images is included.
Similarly, $LG$ and $\overline{LG}$ indicate whether the lens galaxy is present in or absent from the catalog, respectively.
A checkmark (\cmark) and a cross (\xmark) indicate that the corresponding quantity is observationally accessible or not, respectively.
The accuracy of delensing depends not only on the number of accessible observables, particularly the number of detected images, but also on the intrinsic brightness of the source galaxies.
The numbers in parentheses denote the relative error ranges of the recovered magnification factors for a source galaxy of a given brightness.
For the $(G',\overline{LG})$ case, if two or three images of a quadruple-image system are observed, the system is identified as lensed, and delensing is partially possible, though limited.
Conversely, if only one image of a double-image system is observed, the system would not be included in the galaxy-galaxy lensing catalog, as it may be interpreted as a weak lensing system or a special configuration of an unlensed galaxy. 
In this case, delensing is not possible.}
\label{tab:los_terms}
\end{table*}
The integrand of the first term in Eq.~\eqref{LoS_redshift_lensgalaxy}, corresponding to systems that are in-catalog, has a form similar to Eq.~\eqref{eq:complete_catalog} and can be expanded as
\begin{equation}
\begin{aligned}
\label{eq:G'LG_term}
&p\left(z_{s},z_{l},\theta_{l} ,M_{s}, \vec{m}'_{s}\mid GG, \Omega_j, H_{0}, s,l,\mathcal{M}_{\rm SIE}\right)  \\
&= \frac{1}{p\left(s,l \mid  GG, \Omega_j, H_{0} \right)} p\left(z_{s},z_{l},\theta_{l} , M_{s}, \vec{m}'_{s} \mid GG, \Omega_j, H_{0}, \mathcal{M}_{\rm SIE}\right) p\left(s,l \mid z_{s},z_{l},\theta_{l} , M_{s}, \Omega_j, H_{0} ,\mathcal{M}_{\rm SIE}\right) \\
&=  \frac{1}{p\left(s,l \mid  GG, \Omega_j, H_{0} \right)}p(M_{s} | z_{s},z_{l},\theta_{l},\vec{m}'_{s},H_{0},\mathcal{M_{\rm SIE}}) p\left(z_{s},z_{l},\theta_{l}, \vec{m}'_{s} \mid GG, \Omega_j, H_{0}\right) p\left(s,l \mid z_{s},z_{l},\theta_{l}, M_{s}, H_{0} \right),
\end{aligned} 
\end{equation}
where $p(M_{s}|...)$ denotes the conditional probability distribution based on the observed images and the SIE assumption.
In the incomplete catalog case, delensing cannot be assumed fully accurate, and the prior on the intrinsic absolute magnitude of the source galaxy can no longer be represented as a delta function, unlike in the complete catalog case discussed in Sec.~\ref{app:los_complete}.
Nevertheless, the distribution of magnification factors for each source-galaxy image can be inferred through lens reconstruction, using the available information.
This information may vary depending on the possibilities in Table~\ref{tab:los_terms}, allowing the intrinsic absolute magnitude to be recovered to some extent.

Integrating Eq.~\eqref{eq:G'LG_term} over $M_{s}$ and $\vec{m}'_{s}$ yields
\begin{equation}
\begin{aligned}
\label{GG}
&\iint p\left(z_{s},z_{l},\theta_{l} ,M_{s}, \vec{m}'_{s} \mid GG, \Omega_j, H_{0}, s,l, \mathcal{M_{\rm SIE}}\right) d M_{s} d \vec{m}'_{s} \\
&= \frac{1}{p\left(s,l \mid  GG, \Omega_j, H_{0} \right)} \frac{1}{N_{\rm Lgal}(\Omega_{j})}\sum^{N_{\rm Lgal}}_{k} \left[ p\left(z_{s},z_{l}, \theta_{l} \mid \hat{z}_{s,k}, \hat{z}_{l,k} ,\hat{\theta}_{l,k} \right) \left\{ \frac{1}{N_{\rm samp}}\sum^{N_{\rm samp}}_{n} p\left(s,l \mid z_{s},z_{l},\theta_{l}, M^{(n)}_{s}, H_{0} \right)\right\}\right],
\end{aligned}
\end{equation}
where $M^{(n)}_{s}$ denotes the $n^{\rm th}$ sample among the $N_{\rm samp}$ samples drawn from the conditional distribution
$p(M_{s}|...)$, which is generally more informative than a uniform distribution over possible absolute magnitudes of source galaxies.
Furthermore, for the source redshift, even if only a single lensed image is observed, the source redshift ($z_{s}$) can still be determined. 
Therefore, Eq.~\eqref{GG} retains the same form as Eq.~\eqref{incat} with the conditional sampling over $M_{s}$ replacing the delta function.
Again, if luminosity weighting is not applied, retrieval of the true intrinsic luminosity is not required, and the LoS prior reduces to a simpler form.

Next, the integrand of the second term in Eq.~\eqref{LoS_redshift_lensgalaxy}, corresponding to systems that are out-of-catalog, can be expanded in a manner similar to Eq.~(2.11) of \citet{gray2023joint} as follows,
\begin{equation}
\begin{aligned}
\label{eq:out-of-catalog_basic}
& p\left(z_{s},z_{l},\theta_{l} , M_{s}, \vec{m}'_{s} \mid \overline{GG}, \Omega_j, H_{0}, s, l, \mathcal{M}_{\rm SIE}\right) \\
&=\frac{1}{p\left(s, l\mid  \overline{GG}, \Omega_j, H_{0} \right)} p\left(z_{s},z_{l},\theta_{l} , M_{s}, \vec{m}'_{s} \mid \overline{GG}, \Omega_j, H_{0}, \mathcal{M}_{\rm SIE}\right) p\left(s, l \mid z_{s},z_{l},\theta_{l} , M_{s}, \Omega_j, H_{0},\mathcal{M}_{\rm SIE} \right) \\
&= \frac{1}{p\left(s, l \mid \overline{GG}, \Omega_j, H_{0}\right)} \frac{p\left(\overline{GG}\mid z_{s},z_{l},\theta_{l} ,M_{s}, \vec{m}'_{s}, \Omega_j, H_{0}, \mathcal{M}_{\rm SIE}\right) p\left(z_{s},z_{l},\theta_{l}, M_{s}, \vec{m}'_{s} \mid \Omega_j, H_{0}, \mathcal{M}_{\rm SIE}\right)}{p\left(\overline{GG} \mid \Omega_j, H_{0},\mathcal{M}_{\rm SIE} \right)} \\
&\times p(s,l \mid z_{s},z_{l},\theta_{l}, M_{s},H_{0}),
\end{aligned}
\end{equation}
where the term $p\left(\overline{GG} \mid ...\right)$ in the numerator indicates the probability that a galaxy-galaxy lensing system with the specified properties is not included in the catalog.
In the out-of-catalog case, the observability of the lens galaxy is less relevant, and only the source-galaxy redshift is required when constructing the line-of-sight (LoS) redshift prior.
Since the criterion for identifying a system as lensed is the detection of at least two lensed images of the source galaxy, the probability $p\left(\overline{GG} \mid ...\right)$ can be expressed as follows,
\begin{equation}~\label{eq:nested_heavistep}
p\left(\overline{GG}\mid z_{s},z_{l},\theta_{l} ,M_{s}, \vec{m}'_{s}, \Omega_j, H_{0}, \mathcal{M}_{\rm SIE}\right) = \Theta\left(\left[\sum_{k=1}^{N_{\rm img}}\Theta\left(m_{s, k}^{\prime}-m_{\mathrm{th}}(\Omega_{j})\right)\right]-2\right).
\end{equation}
Here, $\Theta(x)$ denotes the Heaviside step function, defined as
$\Theta(x)=1$ for $x\ge 0$ and $\Theta(x)=0$ otherwise.
For the adopted SIE lens model, the total number of images in a strong lensing system, $N_{\rm img}$, can be either 2 or 4, depending on the source position relative to the caustic structure of the lens: sources located inside the caustic produce four images, whereas those outside produce two.

Eq.~\eqref{eq:out-of-catalog_basic} can then be rewritten as
\begin{equation}
\begin{aligned}
\label{eq:out-of-catalog}
&p\left(z_{s},z_{l},\theta_{l} , M_{s}, \vec{m}'_{s} \mid \overline{GG}, \Omega_j, H_{0}, s,l, \mathcal{M}_{\rm SIE}\right)\\
&= \frac{\Theta\left(\left[\sum_{k=1}^{N_{\rm img}}\Theta\left(m_{s, k}^{\prime}-m_{\mathrm{th}}(\Omega_{j})\right)\right]-2\right)}{p\left(s,l \mid \overline{GG}, \Omega_j, H_{0}\right) p\left(\overline{GG} \mid \Omega_j, H_{0} \right)}
\times p\left(z_{s},z_{l},\theta_{l}, M_{s}, \vec{m}'_{s} \mid \Omega_j,H_{0},\mathcal{M}_{\rm SIE}\right)  p(s,l \mid z_{s},z_{l},\theta_{l}, M_{s},H_{0}).
\end{aligned}
\end{equation}
Since the systems under consideration are galaxy-galaxy lensing systems, the lens properties are explicitly included.
The term $p\left(z_{s},z_{l},\theta_{l}, M_{s}, \vec{m}'_{s} \mid  H_{0},\mathcal{M}_{\rm SIE}\right)$ corresponds to the prior distributions of redshifts, lens parameters, and magnitudes, which are independent of the sky pixel $\Omega_{j}$, as the properties of galaxies do not depend on their tangential location.
We adopt minimally informative models to predict galaxy properties since not all relevant galaxy observables are directly accessible.
Unlike the unlensed-catalog case, the observed quantities correspond to the lensed source-galaxy properties and the lens properties.
We assume that the redshifts of source galaxies follow a uniform distribution in comoving volume. 
Accordingly, the redshift distribution of lens galaxies is determined jointly by the source redshift distribution and the lensing optical depth.
For the lens parameters, we adopt the distributions described in Appendix~\ref{app:lens_params}.
Note that both the lensing optical depth and the lens parameters are evaluated under the assumption of SIE profiles, and accordingly, the prior term explicitly depends on $\mathcal{M}_{\rm SIE}$.
Lastly,  we adopt a Schechter function for the intrinsic absolute magnitudes of source galaxies.
Given the source and lens redshifts, the source absolute magnitude, and the lens parameters, the lensed apparent magnitudes of source galaxies can then be computed.

Therefore, the prior can be written as
\begin{equation}~\label{C27}
p\left(z_{s},z_{l},\theta_{l}, M_{s}, \vec{m}'_{s} \mid  H_{0}, \mathcal{M}_{\rm SIE} \right) =\delta(\vec{m}'_{s}-\vec{m}'_{s}(z_{s},z_{l},\theta_{l}, M_{s}, H_{0},\mathcal{M}_{\rm SIE})) p(z_{s},z_{l},\theta_{l}, M_{s} \mid H_{0}).
\end{equation} 

Combining Eq.~\eqref{eq:out-of-catalog} with Eq.~\eqref{C27} and performing the integration over $M_{s}$ and $\vec{m}'_{s}$ yields
\begin{equation}
\begin{aligned}
\label{GGbar}
&\iint  p\left(z_{s},z_{l},\theta_{l} , M_{s}, \vec{m}'_{s} \mid \overline{GG}, \Omega_j, H_{0}, s, l,  \mathcal{M}_{\rm SIE}\right)  d M_{s} d \vec{m}'_{s}\\
&= \frac{1}{p\left(s, l \mid \overline{GG}, \Omega_j, H_{0}\right) p\left(\overline{GG} \mid \Omega_j, H_{0}\right)}
 \int_{M_{s}\left(z_{s},z_{l},\theta_{l}, m_{\mathrm{th}}\left(\Omega_j\right), H_{0}, \mathcal{M}_{\rm SIE}\right)}^{M_{\max }\left(H_0\right)} 
p(z_{s},z_{l},\theta_{l}, M_{s} \mid H_{0}) p(s, l \mid z_{s}, z_{l},\theta_{l}, M_{s}, H_{0}) d M_{s}.
\end{aligned}
\end{equation}
The nested Heaviside step function in Eq.~\eqref{eq:nested_heavistep} has been converted into an integration limit over 
$M_{s}$, which depends on the properties of the source and lens galaxies and is determined by the apparent-magnitude threshold of pixel $j$.

By combining Eqs.~\eqref{GG} and \eqref{GGbar}, the full expression for the LoS redshift in the incomplete galaxy-galaxy lensing catalog case (Eq.~\eqref{LoS_redshift_lensgalaxy}) can be rewritten as
\begin{equation}
\label{full_los}
\begin{aligned}
& p\!\left(z_s, z_l, \theta_l \mid \Omega_j, H_0, s, l ,\mathcal{M}_{\rm SIE}\right) 
\\[-0.5ex]
&\;=\;
\frac{p\left(GG \mid \Omega_j, H_0, s,l,\mathcal{M}_{\rm SIE} \right)}
     {p\left(s,l \mid GG, \Omega_j, H_{0} \right)}
\frac{1}{N_{\rm Lgal}(\Omega_{j})}
\sum_{k}^{N_{\rm Lgal}}p\left(z_{s},z_{l}, \theta_{l} \mid \hat{z}_{s,k}, \hat{z}_{l,k} ,\hat{\theta}_{l,k} \right)
\left[
\frac{1}{N_{\rm samp}}\sum_{n}^{N_{\rm samp}}
p\left(s,l \mid z_{s},z_{l},\theta_{l}, M^{(n)}_{s}, H_{0} \right)
\right]
\\[0.8ex]
&\quad+
\frac{p\left(\overline{GG} \mid \Omega_j, H_0, s,l ,\mathcal{M}_{\rm SIE}\right)}
     {p\left(s,l \mid \overline{GG}, \Omega_j, H_{0}\right)
      p\left(\overline{GG} \mid \Omega_j, H_{0}\right)}
\int_{M_{s}\!\left(z_{s},z_{l},\theta_{l},
m_{\mathrm{th}}\!\left(\Omega_j\right),
H_{0}, \mathcal{M}_{\rm SIE}\right)}^{M_{\max }\left(H_0\right)}
p(z_{s},z_{l},\theta_{l}, M_{s} \mid H_{0})
\, p(s,l \mid z_{s},z_{l},\theta_{l}, M_{s},H_{0}) \, d M_{s} \,.
\end{aligned}
\end{equation}
By applying Bayes' theorem to $p(g|\Omega_{j},H_{0},s)$ terms in each in- and out-of-catalog term, Eq.~\eqref{full_los} can be simplified as,
\begin{equation}
\begin{aligned}
\label{full_los_simplified}
&p\left(z_s, z_l, \theta_l \mid \Omega_j, H_0, s,l,\mathcal{M}_{\rm SIE} \right) \\
&= \frac{1}{p(s,l \mid \Omega_{j},H_{0})} \left[ \frac{p(GG \mid \Omega_{j}, H_{0},\mathcal{M}_{\rm SIE})}{N_{\rm Lgal}(\Omega_{j})}\sum^{N_{\rm Lgal}}_{k} \left\{ p\left(z_{s},z_{l}, \theta_{l} \mid \hat{z}_{s,k}, \hat{z}_{l,k} ,\hat{\theta}_{l,k} \right) \left( \frac{1}{N_{\rm samp}}\sum^{N_{\rm samp}}_{n} p\left(s,l \mid z_{s},z_{l},\theta_{l}, M^{(n)}_{s}, H_{0}  \right)\right) \right\} \right. \\
&\left. +\int_{M_{s}\left(z_{s},z_{l},\theta_{l}, m_{\mathrm{th}}\left(\Omega_j\right), H_{0}, \mathcal{M}_{\rm SIE}\right)}^{M_{\max }\left(H_0\right)} 
p(z_{s},z_{l},\theta_{l}, M_{s} \mid H_{0}) p(s,l \mid z_{s},z_{l},\theta_{l}, M_{s}, H_{0}) d M_{s} \right],
\end{aligned}
\end{equation}
where $p(GG \mid \Omega_{j}, H_{0},\mathcal{M}_{\rm SIE})$ denotes the probability that a galaxy-galaxy lensing system located on the $j^{\rm th}$ pixel is in-catalog.
It is important to note that this expression does not depend on $s$, so we must consider not only the host galaxies but also all other galaxies within the given redshift cut that satisfy the magnitude threshold.
The explicit definition of this probability can be written as
\begin{equation}
\begin{aligned}
&p\left(GG \mid \Omega_j, H_{0},\mathcal{M}_{\rm SIE}\right)\\
&=\iiiint \! \! \!\int p\left(GG \mid z_{s},z_{l},\theta_{l}, M_{s}, \vec{m}'_{s}, \Omega_j, H_{0},\mathcal{M}_{\rm SIE}\right) p\left(z_{s},z_{l},\theta_{l}, M_{s}, \vec{m}'_{s} \mid \Omega_j, H_{0},\mathcal{M}_{\rm SIE}\right)  d M_{s} d \vec{m}'_{s} d z_{s} d z_l \, d\theta_l  \\
&=\iiiint \! \! \!\int\Theta\left(2-\left[\sum_{k=1}^{N_{\rm img}}\Theta\left(m_{s, k}^{\prime}-m_{\mathrm{th}}(\Omega_{j})\right)\right]\right) p\left(z_{s},z_{l},\theta_{l}, M_{s}, \vec{m}'_{s} \mid \Omega_j , H_{0},\mathcal{M}_{\rm SIE}\right)   d M_{s} d \vec{m}'_{s} d z_{s} d z_l \, d\theta_l\\
&=\iiint p\left(z_{s},z_{l},\theta_{l} \mid H_{0}\right) d z_{s} d z_l \, d\theta_l \int_{M_{\min }\left(H_0\right)}^{M_{s}\left(z_{s},z_{l},\theta_{l}, m_{\mathrm{th}}\left(\Omega_j\right), H_{0}, \mathcal{M}_{\rm SIE}\right)}  p\left( M_{s} \mid H_{0}\right)  d M_{s},
\end{aligned}
\end{equation}
where $p\left(z_{s},z_{l},\theta_{l} \mid H_{0}\right)$ denotes the joint prior on the source and lens redshifts and lens parameters introduced in the out-of-catalog case, while $ p\left( M_{s} \mid H_{0}\right)$ is modeled as a Schechter function describing the intrinsic absolute-magnitude distribution of source galaxies.

\end{document}